\documentclass[12pt]{article}

\usepackage[margin=1 in]{geometry}

\usepackage{latexsym}
\usepackage{amssymb,amsmath, bm}
\usepackage{graphicx}
\usepackage{marvosym}
\usepackage{multirow}
\usepackage{lscape}
\usepackage{float,subcaption}
\usepackage{caption}
\usepackage{hvfloat}
\usepackage[title]{appendix}
\usepackage{longtable,booktabs}
\usepackage{threeparttable}
\usepackage[nodisplayskipstretch]{setspace}

\usepackage{array}
\newcolumntype{C}[1]{>{\centering\let\newline\\\arraybackslash\hspace{0pt}}m{#1}}

\usepackage[markup=underlined]{changes}
\definechangesauthor[color=purple]{MJ}
\definechangesauthor[color=brown]{SC}
\definechangesauthor[color=blue]{Lin}

\usepackage{rotating}

\usepackage{arydshln}

\usepackage[compact]{titlesec}

\providecommand{\keywords}[1]
{
  \small	
  \textbf{Keywords: } #1
}

\usepackage{xr-hyper}

\makeatletter
\newcommand*{\addFileDependency}[1]{
  \typeout{(#1)}
  \@addtofilelist{#1}
  \IfFileExists{#1}{}{\typeout{No file #1.}}
}
\makeatother

\newcommand*{\myexternaldocument}[1]{%
    \externaldocument{#1}%
    \addFileDependency{#1.tex}%
    \addFileDependency{#1.aux}%
}

\myexternaldocument{Appendix}

\usepackage{natbib}

\usepackage{chngcntr}

\usepackage[pdftex, bookmarksopen=true, bookmarksnumbered=true,
pdfstartview=FitH, breaklinks=true, urlbordercolor={0 1 0}, citebordercolor={0 0 1}, hidelinks]{hyperref}

\usepackage{titlesec}

\title{ The promise and perils of point process models of political events}
  \author{Lin Zhu\thanks{Institute of Statistics and Big Data, Renmin University of China, Beijing 100872, China; Department of Statistics, Texas A\&M University, College Station, TX 77843. 
  Email: \href{mailto:linzhu_2017@tamu.edu }{linzhu\_2017@tamu.edu}}
    \hspace{0.75in} Scott J. Cook\thanks{ 
  Department of Political Science, Texas A\&M University, College Station, TX 77843. Email:
      \href{mailto:sjcook@tamu.edu}{sjcook@tamu.edu}} 
\hspace{0.75in}  Mikyoung Jun\thanks{
  Department of Mathematics, University of Houston, Houston, TX 77204. Email:
      \href{mailto:mjun@central.uh.edu }{mjun@central.uh.edu }
    }}

\date{}

\usepackage{verbatim}

\begin{document}

\maketitle
\thispagestyle{empty} 

\begin{abstract}
\noindent Event data are increasingly common in applied political science research. While these data are inherently locational, political scientists predominately analyze aggregate summaries of these events as areal data. In so doing, they lose much of the information inherent to these point pattern data, and much of the flexibility that comes analyzing events using point process models. Recognizing these advantages, applied statisticians have increasingly utilized point process models in the analysis of political events. Yet, this work often neglects inherent limitations of political event data (e.g, geolocation accuracy), which can complicate the direct application of point process models. In this paper, we attempt to bridge this divide: introducing the benefits of point process modeling for political science research, and highlighting the unique challenges political science data pose for these approaches. To ground our discussion, we focus the Global Terrorism Database, using a univariate and bivariate log-Gaussian Cox process model (LGCP) to analyze terror attacks in Nigeria during 2014.  

\end{abstract}


\newpage
\setcounter{page}{1}
\section{Introduction}
\doublespacing
Historically, social scientists interested in group-level behaviors (e.g., strikes, protests, rebellion) have relied on aggregate-level data (e.g., the number of terror attacks in a given country-year). Over the past decade or so, however, there has been a surge in the production and availability of incident-level (or \textit{event}) data. \citet[][p. 548]{schrodt2012} defines an event as a ``discrete incident that can be located at a single time (usually precise to a day) and set of actors,'' thereby distinguishing event data from structural data such as GDP. These event data have facilitated a wealth of sub-national analyses in the social sciences, deepening our understanding of civil conflict, terrorism, social movements, etc.  


However, researchers have yet to fully exploit the potential of event data. While event data contain specific spatial (and temporal) coordinates for each incident, researchers typically locate and aggregate events within areal units (e.g., grid cells, districts, states). That is, researchers analyze spatial point pattern data as if it were instead areal data, since methods for analyzing areal data are more familiar. In so doing, much of the rich incident-level information (e.g., timing, location, number of participants, issue, etc.) is necessarily lost. Not only does this limit the information that researchers can extract from these data, but also forces aggregate-level analysis which may be inappropriate for some research questions. 

\par As an alternative, we argue that researchers should analyze event data in their original form, namely, as point-level (or locational) data. Even in purely descriptive analysis, there are obvious advantages to dealing with points directly. Consider, for example, the amount of information contained in a map of points as opposed to aggregate-level means, with points one easily can discern spatial patterns that are lost or distorted once aggregated. The same issue extends to inferential analysis, where aggregate-level data are well-known to risk bias in analyzing individual-level processes \citep[e.g.,][]{achen1995cross}. This can be easily avoided by using spatial (and spatio-temporal) point process methods \citep{diggle}. In short, point process models aim to describe the point pattern of events. By engaging point-level data directly, these methods allow researchers to retain all of the information in the raw data, flexibly select the spatial scale, account for event-specific covariances and measurement error, and produce continuous predictions over the entire spatial domain. 

Given these advantages, these methods have seen wide use in a variety of research domains, including population ecology, meteorology, epidemiology, etc. Increasingly, these approaches have also seen use by statisticians to analyze political events \citep{python2019bayesian}. However, in much of this research the unique features of social science event data have not been properly recognized. Compared to other point process data (e.g., lightning strikes or earthquakes), social event data are more inaccurate \citep{weidmann2015accuracy} or incomplete \citep{cook2017two}. These limitations complicate the direct application of point process methods. In the following, we attempt to bridge this divide: introducing the benefits of point process modeling for political science research, and highlighting the unique challenges of political science event data for these models. 

To ground our discussion, we focus the Global Terrorism Database (GTD, \cite{start}), using a univariate and bivariate log-Gaussian Cox process model to analyze terror attacks in Nigeria during 2014. Among other things, our analysis demonstrates the importance of both analyzing these data as spatial points instead of area frequencies and separately accounting for attacks by individual groups. Failing to do so masks interesting spatial patterns in the data due to aggregation over space or across groups, which limits our overall understanding of subnational terrorism dynamics. Throughout, we also demonstrate several areas where point process models can be generalized to better suit event data. 


\section{Spatial data}\label{section:spatial_data}

Researchers in political science frequently analyze spatial data, by which we mean any data that is geographically referenced. Most observational (i.e., non-experimental) data are inherently spatial, as these data are collected to better understand activities in (or by) spatially defined units (e.g., city, county, country). While political scientists have become increasingly attentive to modeling spatial (inter)dependence \citep[e.g.,][]{beck2006space, franzese2007spatial} in standard regression models, many aspects of spatial statistics remain unfamiliar. 

There are three commonly-used types of spatial data: i) point-referenced (or geostatistical) data, ii) point pattern (or point process) data, and iii) areal (or lattice) data. The first, \emph{point-referenced} data, are continuously-varying data that are collected at a finite (and known) number of locations. Point-referenced data would include, for example, temperature, rainfall, and pollutant concentration, etc. With point-referenced data, the main empirical challenge is drawing inferences over the entire continuously-varying spatial domain from a discrete set of observations. These types of data are relatively rare in political science, so we do not devote much attention to them here.\footnote{Notable exceptions include \cite{tam2007prospecting}, which uses geostatitical methods to analyze campaign donations, and \cite{FreemanEtAl} which uses geostatitical methods to validate event geolocations from machine-coded data.}

\begin{figure}[hbt!]
\centering
\includegraphics[width=0.6\textwidth]{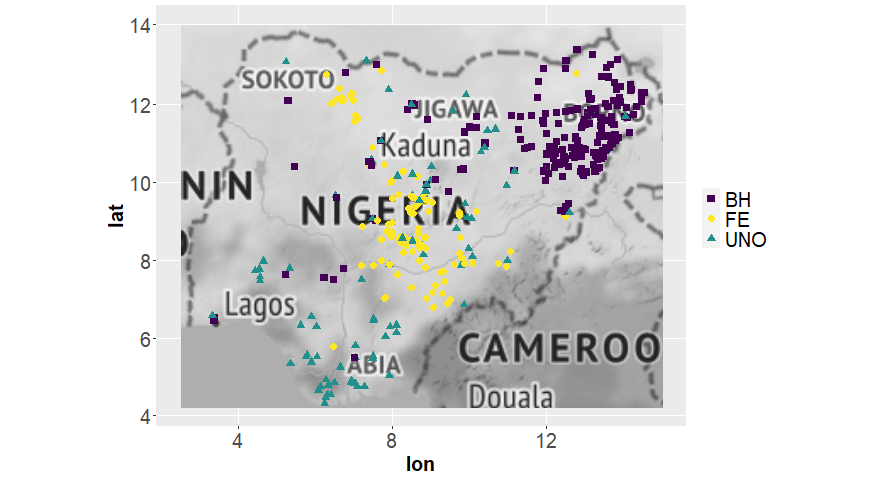}
\caption{2014 terror attacks by Boko Haram (BH), Fulani extremists (FE), and others (UNO).}
\label{fig:pointpattern}
\end{figure}

\textit{Point pattern} data, on the other hand, are increasingly common in political science. As in point-referenced data, one assumes a continuous process that is only observed for a finite set of locations, however, in point pattern data these \emph{locations} are of theoretical interest. Also called ``presence only" data, with point pattern data our observations are the points themselves (e.g., potholes, protests, shootings, etc.), which indicate the occurrence of some event at a given location. Figure~\ref{fig:pointpattern} offers an example, plotting the locations of terror attacks in Nigeria during 2014. 

Finally, \textit{Areal} data is by far the most widely used type of spatial data in political science. Here the spatial domain is partitioned into a pre-determined set of regular (e.g., PRIO grid cells) or irregular (e.g., counties) lattices, and variable(s) are defined for those areal units. Areal data analysis is widely used for three reasons. First, researchers are often explicitly interested in the actions, behaviors, and outcomes of (areal) spatial units (e.g., districts, counties, states). Second, data are often only collected and made available as aggregate summary values (e.g., unemployment, population) for a given spatial unit. Lastly, areal data is produced because the researchers themselves have chosen to aggregate point pattern data into areal units \citep{ward2018spatial}.

It is this final practice -- treating point pattern data as areal data -- which motivates our discussion. As discussed in  Section~\ref{section:point_process_model}, there are several unique advantages to applying spatial point process models to point pattern data, however, there is rarely a strong rationale for preferring (pre-)aggregation of point pattern data by default. 

\section{Spatial point process models}\label{section:point_process_model}

Spatial point pattern data are comprised of a random number of events occurring at random locations in a spatial region. Each event $i$ is defined by its location, $\mathbf{y}_i$, with $n$ events recorded as $\mathbf{y}=\{\mathbf{y}_1,\ldots,\mathbf{y}_n\}$. Often each $i^{th}$ event is denoted using latitude (lat) and longitude (lon), such that  $\mathbf{y}=\{\text{(lat,lon)}_1,\ldots,\text{(lat,lon)}_n\}$. These observed data are modeled as a {\it spatial point process} and the underlying stochastic model treats both spatial locations ($\mathbf{y}_i$) and the number of events ($n$) as random \citep{diggle}. In analyzing point pattern data, the assumed spatial point process and model is selected by the researcher. In the following, we detail two commonly used spatial point process models: 1) the Poisson process, as this is the building block of many spatial point process models; and 2) the Log-Gaussian Cox process (LGCP) model, a more flexible generalization of the Poisson process model better suited for most political science data. 

\subsection{Poisson processes}\label{sec:Poisson_process}

To fix terms, we consider a point process $Y=\{Y(\mathbf{s}):\mathbf{s} \in D\}$ defined in a spatial domain $D \subset \mathbb{R}^d$ (commonly, $d=2)$. Let $N(A):=N_Y(A)$ denote the number of points or events that $Y$ places in a bounded spatial region $A \subset D$. The point process $Y$ is called {\it Poisson process} if $N(A)$ follows a Poisson distribution with the rate, 
\begin{align}
    \mu(A) = \int_A \lambda(\mathbf{s})\text{d}\mathbf{s}. \label{eq:rate}
    \end{align} 
    Here, $\lambda(\mathbf{s})$ is the {\it first-order intensity function} of $Y$,
\begin{align}\label{lambda(s)}
    \lambda(\mathbf{s})=\lim_{|\text{d}\mathbf{s}|\rightarrow 0}\frac{\mathbb{E}\{N(\text{d}\mathbf{s})\}}{|\text{d}\mathbf{s}|} \quad \mathbf{s}\in D,
\end{align}
where $\text{d}\mathbf{s}$ is the (infinitesimal) spatial region that contains point $\mathbf{s} \in D$, $|\text{d}\mathbf{s}|$ is the area of $\text{d}\mathbf{s}$, and $\mathbb{E}\{N(\text{d}\mathbf{s})\}$ represents the expectation of the random variable $N(\text{d}\mathbf{s})$. The mapping $\mathbf{s} \mapsto\lambda(\mathbf{s})$ describes the spatial mean structure of Poisson distribution, with the first-order intensity function $\lambda(\mathbf{s})$ giving the expected count for a given spatial location. When $\lambda(\mathbf{s})$ does not vary by location, i.e. $\lambda(\mathbf{s}) = \lambda$ for all $\mathbf{s}$, the point process is considered {\it homogeneous}, as it has a constant rate over the spatial domain. 


In most real-world applications, however, a constant first-order intensity function is too restrictive. For example, in the Nigerian terrorism data in Figure~\ref{fig:pointpattern} the attack pattern is not evenly distributed across the country. Instead, it appears that a spatially varying intensity function -- indicating an {\it inhomogeneous} spatial point process -- may be more appropriate. For inhomogeneous processes, it is common to express the spatial intensity function as a linear combination of covariates:  
\begin{align}
    f\{\lambda(\mathbf{s})\} = \beta_0 +\beta_1 V_1(\mathbf{s}) +\cdots+\beta_p V_p(\mathbf{s}), \label{eq:lambda(s)}
\end{align}
where $f(\cdot)$ is a reversible transform (often an identify function or log transformation),   $V_1(\mathbf{s}),\ldots,V_p(\mathbf{s})$ are point-referenced explanatory variables defined at $\mathbf{s}$, and $\beta_i$'s are unknown coefficients. 

Even with a more fully specified first-order intensity function, researchers will often want to consider the second-order properties of these point process models. Second-order structure (i.e., covariance) in point process models characterizes ``interactions'' between points, that is, spatial dependence. If the points are spatially independent of one another, then a point process satisfies {\it Complete Spatial Randomness (CSR)}. Under CSR, the number of points in spatially disjoint regions are unrelated and follow the Poisson distribution with an intensity function \eqref{eq:lambda(s)}. This is unlikely in real-world applications, however, as we often observe evidence of spatial interactions among events (i.e. points), such as spatial clustering and repulsiveness. These potential interactions can be assessed via second-order quantities of point processes (analogous to the covariance function of Gaussian processes). Specifically, the {\it second-order intensity} is defined as 
\begin{align}\label{eq:lambda^(2)}
    \lambda^{(2)}(\mathbf{s},\mathbf{u})=\lim_{|\text{d}\mathbf{s}|,|\text{d}\mathbf{u}|\rightarrow 0} \frac{\mathbb{E}\{N(\text{d}\mathbf{s})N(\text{d}\mathbf{u})\}}{|\text{d}\mathbf{s}||\text{d}\mathbf{u}|},
\end{align}
where
$\text{d}\mathbf{s}$ and $\text{d}\mathbf{u}$ respectively represent
the infinitesimal spatial regions that contain points $\mathbf{s}$ and $\mathbf{u}$. Informally, this reflects the probability of any pair of events occurring in the vicinity of $\bf s$ and $\bf u$ \citep{bivand2013applied}. More specifically, the quantity  $\lim_{|\text{d}\mathbf{s}|\rightarrow 0} N(\text{d}\mathbf{s})/|\text{d}\mathbf{s}|$ has a covariance function  $\lambda^{(2)}(\mathbf{s},\mathbf{u})-\lambda(\mathbf{s})\lambda(\mathbf{u})$, which is called the {\it covariance density} of the point process. Generally speaking, positive and negative covariance densities correspond to concentration and dispersion, respectively. Under CSR, the second-order intensity \eqref{eq:lambda^(2)} reduces to $\lambda(\mathbf{s}) \lambda(\mathbf{u})$ because $N(\text{d}\mathbf{s})$ and $N(\text{d}\mathbf{u})$ are independent for $\mathbf{s} \neq \mathbf{u} \in D$. When CSR is not satisfied, researchers should explicitly model the covariance, as failing to do so risks inaccurate estimates of model parameters, which diminishes the ability of researchers to make credible inferences or predictions. 

\subsection{Ripley's K function}
\label{sec:prelim}

One commonly used diagnostic to assess spatial randomness in spatial point pattern data is Ripley's K function \citep{ripley1976second}. As with many tests of dependence (e.g., $\chi^2$, Moran's I, etc.), Ripley's K compares the observed spatial pattern in the sample (hereafter the empirical K function) with what would be expected under spatial randomness (hereafter the theoretical K function). 

The theoretical K function can be calculated for homogeneous or inhomogenous point processes, however, we focus on inhomogenous point processes as they are the more general case. \citet[][p. 332]{baddeley2000non} give the theoretical inhomogenous K function as:
\begin{align}\label{eq:Kinhom(r)}
    K_{\text{inhom}}(r)=\mathbb{E} \Bigl\{ \sum_{\mathbf{s} \in Y}{\frac{1}{\lambda(\mathbf{s})} \mathbf{1}_{ \{0< \|\mathbf{s}-\mathbf{u} \| \leq r\}}}\Bigr \}
    ,~~ r \geq 0, 
\end{align}  
where $\mathbf{s}$ and $\mathbf{u}$ are random points of $Y$ contained with the spatial domain $D$. For a given set of points $\mathbf{s}$ and $\mathbf{u}$, we evaluate whether the distance between them (i.e., $\|\mathbf{s}-\mathbf{u} \| $) is less than some $r$. That is, whether the points are sufficiently close to one another. If so, then the indicator function $\mathbf{1}_{\{\cdot \}}$ assumes a value of 1, and, if not, it produces a zero. Repeating this for all ${\mathbf{s} \in Y}$ indicates the number of ``neighboring'' events within some distance $r$ of an arbitrarily selected event $\mathbf{u}$. This count is normalized by the conditional first-order intensity $\lambda(\mathbf{s})$, as a greater intensity produces more events regardless of the spatial dependence.\footnote{For homogeneous point processes $K(r)$ is calculated as in \eqref{eq:Kinhom(r)} with a common $\lambda$ substituted in for $\lambda(\mathbf{s})$.} 

We then compare this to the empirical inhomogeneous $K$ function, which is obtained from 
\begin{align}\label{eq:estimate_Kinhom}
    \hat{K}_{\text{inhom}}(r)=\sum_{\mathbf{s} \in Y}\sum_{\mathbf{u} \in Y}\frac{c(\mathbf{s},\mathbf{u},r)\textbf{1}_{\{0<\|\mathbf{s}-\mathbf{u}\|\leq r\}}}{\hat{\lambda}(\mathbf{s})\hat{\lambda}(\mathbf{u})}. 
\end{align}
Where $\mathbf{s}$ and $\mathbf{u}$ are again points in $Y$, $\hat{\lambda}(\cdot)$'s are estimated first-order intensities, and $c(\mathbf{s},\mathbf{u},r)$ is an edge correction factor. In an observed point pattern, the neighbors of points at the edge (i.e., near the border) are omitted when outside the observational window. This can bias ``neighboring'' counts for these points, and consequently the sample-wide measure of dependence. As such, corrections for these edge effects are often employed (see  \cite{cressie2015statistics}). Here, we use the Ripley's isotropic correction,  $c(\mathbf{s},\mathbf{u},r) = \{|D| \cdot g(\mathbf{s},\mathbf{u})\}^{-1}$, where $|D|$ is the area of the observational window and $g(\mathbf{s},\mathbf{u})$ is the fraction of the circumference of the circle with centre $\mathbf{s}$ and radius $\|\mathbf{s}-\mathbf{u}\|$ which lies inside the window.

With both the theoretical and empirical K functions calculated, researchers compare the two quantities to determine possible clustering or repulsiveness within the point patterns. Within a radius $r$, $\hat{K}_{\text{inhom}}(r)> K_{\text{inhom}}(r)$ may suggest clustering, and $\hat{K}_{\text{inhom}}(r)<K_{\text{inhom}}(r)$ may suggest dispersion. If $\hat{K}_{\text{inhom}}(r)$ and $K_{\text{inhom}}(r)$ are significantly different, then the CSR assumption is rejected and alternative point process models should be explored. Several methods have been proposed to assess the statistical significance of Ripley's K function. For example, \cite{grabarnik2002goodness} suggest CSR should be rejected if the maximum departure of estimated Ripley's K function from $\pi r^2$ in the original sample exceeds the maximum departure found in $95\%$ of the simulated samples. 


We will demonstrate the utility of Ripley's K for applied research in Section~\ref{sec:detail_estimate_K}, where we also illustrate how it can be extended to assess multivariate spatial point patterns . 
 
\subsection{Log-Gaussian Cox Process models}
Since many spatial point pattern data from real applications do not exhibit the CSR property, it is often necessary to consider more flexible point process models. The {\it Log-Gaussian Cox Process (LGCP)} is a well-developed extension of the Poisson process that can incorporate spatial ``interactions" among points such as spatial clustering \citep{moller_et_al98}. As such, LGCP models have become increasingly popular in numerous fields, including environmental studies, population ecology, finance,  etc. 

\subsubsection{\label{sec:uni_LGCP}Univariate case}\label{sec:univariate}
In an LGCP model, one starts with the Poisson process as defined in Section~\ref{sec:Poisson_process}, however, the intensity function (denoted by $\lambda$ above) is now assumed to be a stochastic Log-Gaussian process (now given as $\Lambda$). Since any realization of the process $\Lambda$ is not constant over the spatial domain, LGCP models naturally produce inhomogeneous Poisson processes. As before, the intensity function is often expressed in terms of covariates as 
\begin{align}
    \log \Lambda(\mathbf{s}) = \beta_0 +\beta_1 V_1(\mathbf{s}) +\cdots+\beta_p V_p(\mathbf{s}) + e(\mathbf{s}), \label{eq:LGCP}
\end{align} where the $V_i(\mathbf{s})$'s are spatially referenced covariates, and $e(\mathbf{s})$ is a Gaussian random variable with mean $\mu$ and variance $\sigma^2$. Since $\exp\{e(\mathbf{s})\}$ is a log-Normal random variable, we have $$ \mathbb{E}[\exp\{e(\mathbf{s})\}]=\exp(\mu+\sigma^2/2).$$ \cite{diggle2013spatial} suggest letting $\mu=-\sigma^2/2$ so that $\mathbb{E}[\exp\{e(\mathbf{s})\}]=1$. This separates covariance parameters of $e$ from the coefficients $\beta_i$, such that $\beta_i$ only affects the first-order intensity, while $\sigma$ affects the second-order intensity. To see this recall \eqref{eq:LGCP} and $\mathbb{E}[\exp\{e(\mathbf{s})\}]=1$, which jointly allow us to show how the the first-order intensity function
\begin{align*}
    \tilde{\lambda}(\mathbf{s}):=\mathbb{E}\{\Lambda(\mathbf{s})\}&=\mathbb{E}[\exp\left\{\beta_0 + \cdots+\beta_p V_p (\mathbf{s}) + e(\mathbf{s}) \right\}] \\
    &= \exp \left\{\beta_0 + \cdots+\beta_p V_p (\mathbf{s}) \right \} \cdot \mathbb{E}[ \exp \left \{e(\mathbf{s})\right \}]\\
    &=\exp \left\{\beta_0 + \cdots+\beta_p V_p (\mathbf{s}) \right \},
\end{align*}
only depends on $\beta_i$'s and the covariates.

For second-order properties, it is common to assume $e(\mathbf{s})$ is stationary and isotropic with a covariance function
\begin{align}
C(h) & :=\text{Cov}\{e(\mathbf{s}),e(\mathbf{u})\}=\sigma^2\rho(h) \quad (\mathbf{s},\mathbf{u}\in D,~ h=\|\mathbf{s}-\mathbf{u}\|),  \label{eq:C(h)}
\end{align}
where $\rho(h)$ is the spatial correlation function.\footnote{Should $\sigma^2=0$ then $e$ in \eqref{eq:LGCP} is a degenerate constant, and it reduces to a Poisson process.}  Throughout this paper, we use an exponential correlation function $\rho(h)=\exp(-{h}/{\phi})$, with a spatial range parameter $\phi$. 

LGCP models allow us to explicitly model interactions among the points via the spatial covariance structure of $\log \Lambda(\mathbf{s})$. The covariance and correlation functions of $\Lambda(\mathbf{s})$, respectively, can be expressed as
\begin{align}
    \text{Cov}\{\Lambda(\mathbf{s}),\Lambda(\mathbf{u})\} &= \mathbb{E}\{\Lambda(\mathbf{s})\Lambda(\mathbf{u})\}-\mathbb{E}\{\Lambda(\mathbf{s})\}\mathbb{E}\{\Lambda(\mathbf{u})\} \nonumber\\
    &= \tilde{\lambda}(\mathbf{s}) \tilde{\lambda}(\mathbf{u}) \left[\exp\{C(h)\}-1 \right],\label{eq:cluster} \\
    \text{Corr}\{\Lambda(\mathbf{s}),\Lambda(\mathbf{
u})\} &= \frac{\exp\{C(h)\}-1}{\exp(\sigma^2)-1}, \label{eq:cluster_corr}
    \end{align}
where $\mathbf{u}$ again represents another random point in X. For $C(h)=\sigma^2\exp(-h/\phi)$, we see that \eqref{eq:cluster} becomes $$\mbox{Cov} \{\Lambda(\mathbf{s}), \Lambda(\mathbf{u})\} \propto \exp{\{\sigma^2 \exp(-h/\phi)\}}-1,$$ 
which is a monotonically decreasing function of distance lag $h$, determined by $\sigma^2$ and $\phi$.

Overall, spatial clusters are related to both the spatial range parameter, $\phi$, and the variance parameter, $\sigma^2$. This can be seen directly in equations \eqref{eq:C(h)} through \eqref{eq:cluster_corr}. As either of these paramaters tends to zero, then $\mbox{Cov}\{ \Lambda(\mathbf{s}),\Lambda(\mathbf{u})\}\approx 0$ and there is no spatial clustering. Increasing $\phi$, while holding $\sigma^2$ fixed (and non-zero), determines the size (i.e., radius) of the clusters. Instead, if one holds $\phi$ fixed (and non-zero), then increasing $\sigma^2$ determines the variance in the intensity across the spatial domain. Jointly increasing both $\phi$ and $\sigma^2$ directly affects the dependence of process $\Lambda$, producing larger clusters of points (due mainly to $\phi$) that are increasingly distinct from those points not in a given cluster (due mainly to $\sigma^2$). In Appendix \ref{LGCP_cov_sims} we use simulations to visually demonstrate the spatial patterns produced under different values of $\phi$ and $\sigma^2$.

\subsubsection{\label{sec:multi}Multivariate case}

With two or more potentially related spatial point patterns (e.g., terrorist attack data from different groups), one needs multivariate point process models. These multivariate models account for the marginal spatial point process (as in the univariate case) but importantly also account for cross-group interactions. 

Extending the univariate LGCP models discussed in Section~\ref{sec:univariate}, consider $(Y_1,\ldots, Y_q)$ as $q$ spatial point patterns with a stochastic intensity $\Lambda=(\Lambda_1,\cdots,\Lambda_q)$. We express the intensity in terms of covariates and employ a multivariate spatial covariance function to model the spatial dependence of $\mathbf{e} = (e_1, \ldots, e_q)$. This reflects not only marginal processes but also any interaction between pairs of spatial point patterns. For example, in a bivariate LGCP model the first-order intensity is given as 
\begin{align} \tilde{\lambda}_j(\mathbf{s}):=\mathbb{E}\{\Lambda_j(\mathbf{s})\}=\exp\{\beta_{0,j}+\cdots+\beta_{p,j}V_p(\mathbf{s})\}, ~~ (j=1,2).\label{constraint}
\end{align}
We then define the cross-covariance function between $e_1$ and $e_2$ as 
\begin{align}
    C_{12}(h)&:=\mbox{Cov}\{e_1(\mathbf{s}),e_2(\mathbf{u})\} = \sigma_1\sigma_2 \rho_{12}(h),
    \label{12}
\end{align}
where $\sigma_j^2=\text{Var}\{e_j(\mathbf{s})\}$, $h=\|\mathbf{s}-\mathbf{u}\|$, and $\rho_{12}(h)$ is the cross-correlation function between $e_1$ and $e_2$.\footnote{As in the univariate case, $e_j(\mathbf{s})$ is a Gaussian random variable with mean $-\sigma_j^2/2$ and variance $\sigma_j^2$.} Analogous to \eqref{eq:cluster} and \eqref{eq:cluster_corr}, the induced spatial cross-covariance and cross-correlation function between $\Lambda_1(\mathbf{s})$ and $\Lambda_2(\mathbf{u})$ are written as
\begin{align}\label{eq:Cov_Lambda_12}
    \text{Cov}\{\Lambda_1(\mathbf{s}),\Lambda_2(\mathbf{
u})\} &=\tilde{\lambda}_1(\mathbf{s})\tilde{\lambda}_2(\mathbf{u})[\exp\{C_{12}(h)\}-1],\\
\text{Corr}\{\Lambda_1(\mathbf{s}),\Lambda_2(\mathbf{
u})\} &= \frac{\exp\{C_{12}(h)\}-1}{\sqrt{\exp(\sigma_1^2)-1}\sqrt{\exp(\sigma_2^2)-1}}. \label{13}
\end{align}
From \eqref{12} and \eqref{13}, we see that cross-dependence between pairs of point patterns is related to variance parameters of two intensity processes ($\sigma_1^2$ and $\sigma_2^2$) and the cross-correlation function, $\rho_{12}(h)$. The form of $\rho_{12}(h)$ is determined by the assumed cross-covariance structure of $\Lambda_1$ and $\Lambda_2$, as we discuss in the next paragraph. In general, however, if $\rho_{12}(h) \approx 0$, then $\text{Cov}\{\Lambda_1(\mathbf{s}),\Lambda_2(\mathbf{
u})\} \approx 0$, indicating that two point patterns are unrelated. Alternatively, a positively (negatively) signed $\rho_{12}(h)$ implies a positive (negative) $\text{Cov}\{\Lambda_1(\mathbf{s}),\Lambda_2(\mathbf{
u})\}$ and hence clustering (separation) of pairs of point patterns. 
 
As noted, estimation of these parameters depends on the particular cross-covariance function employed. The development of these cross-covariance functions for multivariate Gaussian processes is an active area in spatial statistics (see \cite{genton_kleiber15}). In our application below we use the {\it Linear Model of Coregionalization (LMC)}, which exploits the fact that each spatial process can be written as a linear combination of individual and common spatial processes \citep{gelfand_et_al04}. An \textsf{R} package by \cite{taylor2015bayesian}, {\it lgcp}, uses this approach, decomposing $e_1$ and $e_2$ as 
\begin{align} 
e_1= W_1 + W, ~~ \mbox{and}~~ e_2 =  W_2 + W,\label{LMC}
\end{align} where $W_1$, $W_2$, and $W$ are each independent Gaussian processes with variances $\sigma_{W_1}^2$, $\sigma_{W_2}^2$, and $\sigma_W^2$, respectively. Although $W_1$, $W_2$, and $W$ are independent, $e_1$ and $e_2$ have non-zero spatial covariance through the ``common" process $W$. As in the univariate case, we use an exponential covariance functions for $W_j$'s and $W$:
\begin{align*}
    \mbox{Cov}\{W_j(\mathbf{s}), W_j(\mathbf{u})\} &= \sigma_{W_j}^2 \exp{(-\|\mathbf{s}-\mathbf{u}\|/\phi_{W_j})},\\ \mbox{Cov}\{W(\mathbf{s}), W(\mathbf{u})\} &= \sigma_W^2 \exp{(-\|\mathbf{s}-\mathbf{u}\|/\phi_W)},
\end{align*}
where $\phi_{W_j}$'s and $\phi_W$, are spatial range parameters for $W_j$'s and $W$, respectively. Under the LMC structure, the size of spatial clustering is determined by all three spatial range parameters in a bivariate LGCP model: $\phi_{W_1}$, $\phi_{W_2}$, and $\phi_W$. The interaction across the two point patterns -- i.e., size of cross-clustering -- is mainly determined by $W$ as given in \eqref{eq:cov_eij}. When $\phi_W \approx 0$, then $\rho_{12}(h) \approx 0$ and the two point patterns will be approximately unrelated. Conversely, when $\phi_W$ is large -- and $\sigma^2_W$ is large relative to $\sigma^2_{W_1}$ and $\sigma^2_{W_2}$ --  the two point patterns may exhibit strong repulsion. 

While we prefer this approach, there are drawbacks to the LMC covariance structure. First, since each $e_j$ is a combination of $W_j$ \emph{and} $W$, $\phi_W$ not only affects the size of cross-clustering but also the size of marginal spatial clustering. As such, the common process affects both the marginal and joint structure, complicating interpretation of spatial covariance parameters. Second, the specific implementation of the LMC structure given in \eqref{LMC} requires non-negative cross-covariance. This is a significant limitation, as many bivariate applications in political science are theoretically expected to exhibit negative cross-group correlations (e.g., free-riding dynamics, late-mover advantages, etc.). As such, we take on this issue directly in Section \ref{sec:cov_info}, modifying the LMC structure to permit negative cross-covariance. 

\subsection{Summary}
In our brief survey of spatial point pattern models, we identified several potential advantages of point process methods for the analysis of presence only data: i) inherits the original data structure, ii) retains all information contained in the raw data (as opposed to sums/averages), iii)  flexible selection of spatial units/scale (as opposed to pre-selected, fixed units), iv) can account for event-specific covariances, v) can account for event-specific measurement error, vi) straight-forward multivariate extensions, vii) offers continuous predictions over the entire spatial domain. Why, then, have social scientists lagged in the adoption of these methods? We believe it may either be due to a lack of familiarity with these methods or a belief that political science data are ill-suited for these approaches, as such we take on each of these possibilities in our applied example in the next section. 

\section{Application: Terrorism in Nigeria}\label{section:applications}

In the following, we analyze data on terrorism in Nigeria during 2014 using the spatial point process methods discussed above. First, we use Ripley's K and cross-K function to assess the spatial distribution of these attacks. Second, we fit both univariate and bivariate LGCP models to attacks from Boko Haram (BH) and Fulani Extremists (FE). Finally, we discuss several limitations of political events data, show the consequences of these in our illustration, and anticipate areas for further development in spatial point process models. 
\subsection{\label{sec:detail_estimate_K}Preliminary analysis with Ripley's K function }

A visual assessment of Figure \ref{fig:pointpattern} suggests that the intensity of terror attacks across Nigeria is inhomogeneous, with potential marginal and joint interaction between points. Rather than rely on visual inspection alone, we estimate the inhomogeneous K and cross-K function to assess potential spatial clustering (or repulsiveness) within and across terror attacks. First, we calculate the empirical inhomogeneous K functions for each observed spatial pattern separately and cross-K functions of two groups (BH and FE).\footnote{The empirical K functions are achieved using the {\it spatstat} package in \textsf{R}, where intensity estimates of data points are calculated via a kernel-smoothed function. For example, $\hat{\lambda}(\mathbf{y}_i)=\sum_{j \neq i}k(\mathbf{y}_j-\mathbf{y}_i)w(\mathbf{y}_j) / \sum_{j \neq i}k(\mathbf{y}_j-\mathbf{y}_i)$ is the weighted average of kernel contributions from other data points, where $k()$ is the smoothing Gaussian kernel and $w()$ is the weight of data point masses.} Second, using the kernel-smoothed intensity functions, we simulate 1000 inhomogeneous Poisson point patterns of BH attacks, FE attacks, and All attacks. From these simulated results, we construct a 95\% confidence envelope, which allows us to compare the spatial pattern found in the observed data to what would be expected under complete spatial randomness. 

\begin{figure}[hbt!]
      \begin{subfigure}{0.32\textwidth} 
      \centering
    \includegraphics[width=0.99\linewidth]{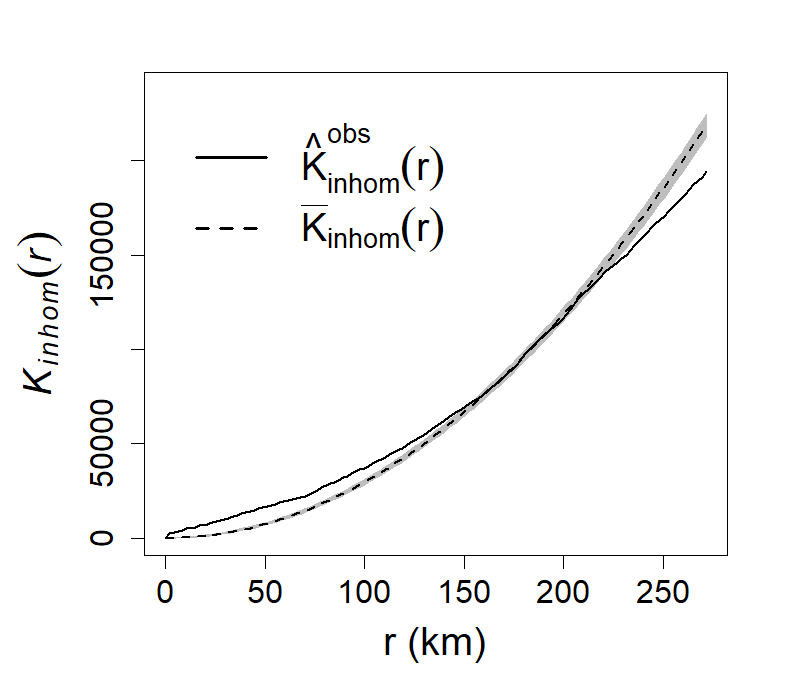}
    \caption{All data}
    \label{fig:Kenv_all}
    \end{subfigure}
    \vspace{-4mm}
          \begin{subfigure}{0.32\textwidth} 
                \centering
    \includegraphics[width=0.99\linewidth]{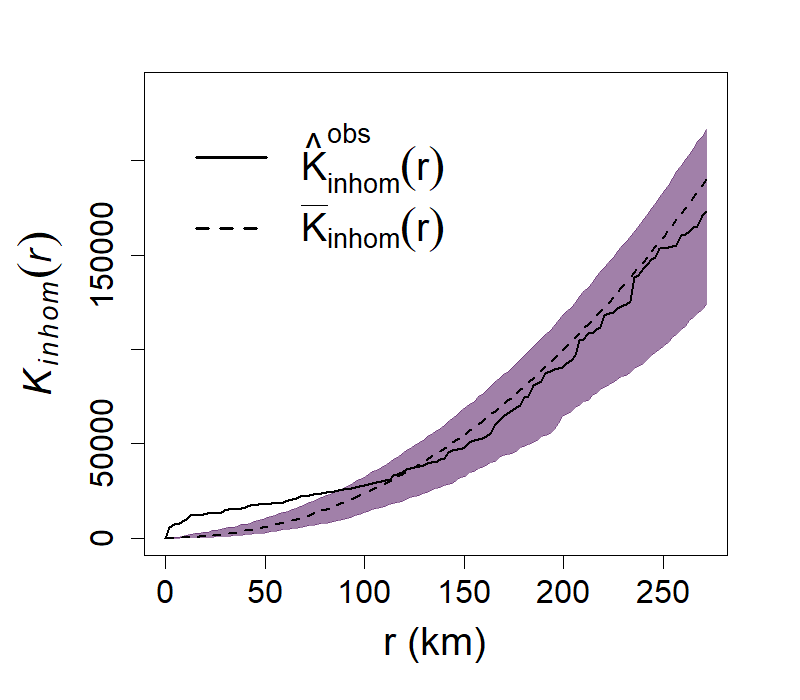}
    \caption{BH}
    \label{fig:Kenv_BH}
    \end{subfigure}
 \begin{subfigure}{0.32\textwidth}
       \centering
 \includegraphics[width=0.99\linewidth]{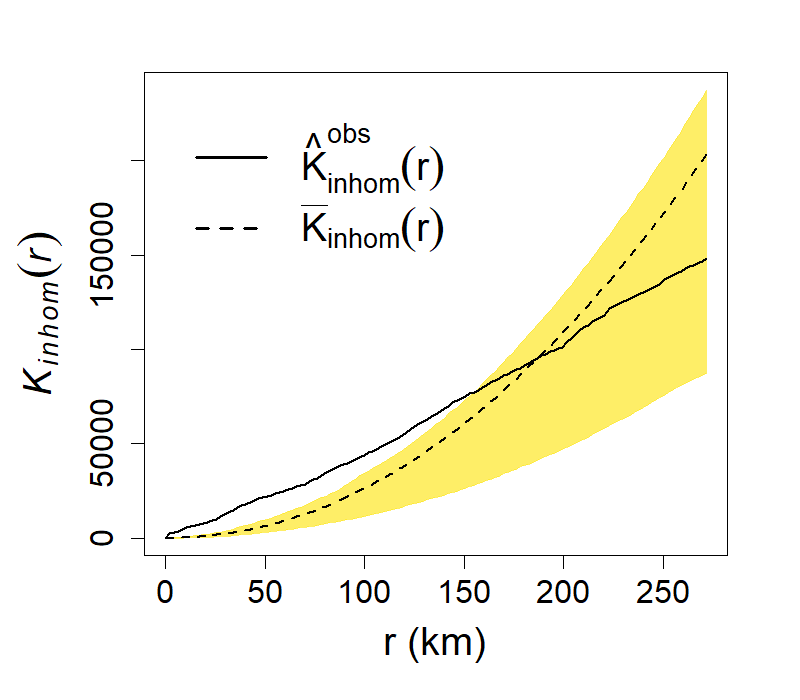}
    \caption{FE}
    \label{fig:Kenv_FE}
    \end{subfigure}
   \caption{Empirical inhomogeneous K (solid) and inhomogeneous K under CSR (dashed) with 95\% confidence envelope.}
    \label{fig:Kenv2} 
\end{figure}

The results are given in Figure \ref{fig:Kenv2} with each panel providing 3 quantities: i) the empirical inhomogemeous K value, $\hat{K}^{\text{obs}}_{\text{inhom}}(r)$, ii) the mean of the simulated inhomogemeous K values under spatial randomness, $\bar{K}_{\text{inhom}}(r)$, and iii) the 95\% envelope for the simulated K values. The top-left panel, Figure \ref{fig:Kenv_all}, reports these K values when all terror attacks in Nigeria are included. We observe that at distances around $r<150~\text{km}$ the estimated K function of the observed point pattern (the solid black line) is greater than the upper bound of 95\% confidence region of the simulated data, thereby indicating spatial clustering of points up to approximately 150 km. This implies that the conditional independence assumed under the spatial Poisson process is not supported by these data. 

Similar findings obtain when we break the attacks out by the perpetrator (BH or FE) in Figures \ref{fig:Kenv_BH} and \ref{fig:Kenv_FE}, however, the two groups show distinct cluster sizes. For BH (Figure \ref{fig:Kenv_BH}), the empirical K function values are greater than the 95\% confidence region up to approximately 80 km, whereas for FE (Figure \ref{fig:Kenv_FE}) this distance is roughly 150 km. These differences in relative cluster sizes are consistent with what we observe in Figure \ref{fig:pointpattern}, with attacks by FE exhibiting a larger cluster pattern. The width of the 95\% confidence regions in Figures \ref{fig:Kenv_all}, \ref{fig:Kenv_BH}, and \ref{fig:Kenv_FE} is also consistent with our understanding of the data, as there are the most attacks in the ``all data'' sample (i.e., Figure \ref{fig:Kenv_all}) and the fewest in the FE sample (i.e., Figure \ref{fig:Kenv_FE}).\footnote{In 2014, there 714 total attacks in Nigera, including 436 by BH and 156 by FE.} As before, \ref{fig:Kenv_BH}, and \ref{fig:Kenv_FE} implies that usual Poisson process models will not be adequate for these data due to apparent spatial clustering.
\begin{figure}[hbt!]
\centering
        \includegraphics[width=0.4\linewidth]{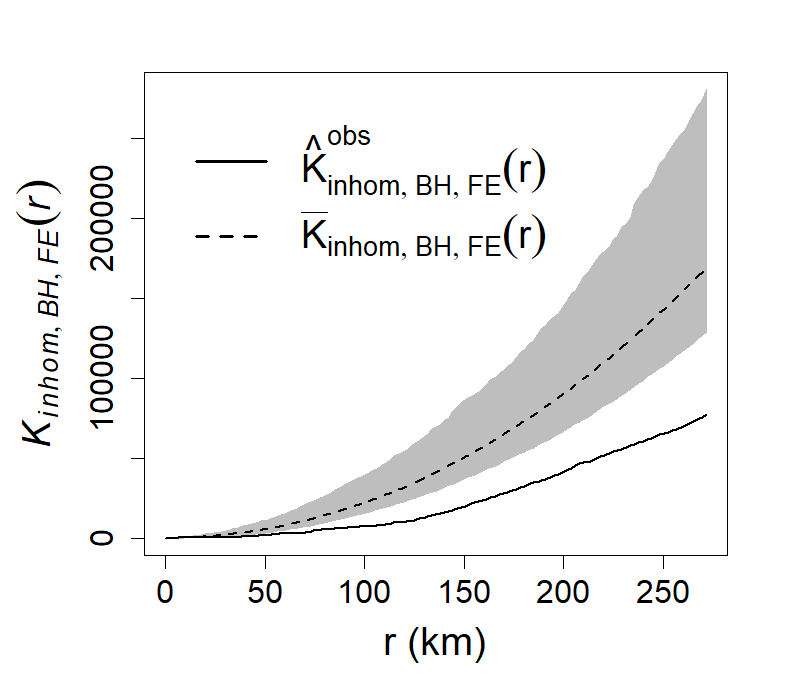}
   \caption{ Empirical inhomogeneous K (solid) and inhomogeneous K under CSR (dashed) with 95\% confidence envelope. }
       \label{fig:crossK_env_mix}
\end{figure}

We repeat this process for the bivariate analysis: estimate the empirical cross-K function for BH attacks and FE attacks, and then simulate 1000 independent bivariate inhomogeneous Poisson point patterns using the marginal kernel-smoothed intensity functions. This allows us to assess whether there is dependence \emph{across} different point patterns. The results of the cross-K function analysis are given in Figure~\ref{fig:crossK_env_mix}, where $\hat{K}^{\text{obs}}_{\text{inhom},BH,FE}(r)$ is the empirical inhomogeneous cross-K value of two point patterns (solid line), $\bar{K}_{\text{inhom},BH,FE}(r)$ is the mean of the simulated cross-K functions assuming independence (dashed line), and the shaded area is the 95\% envelope. Here we observe that the empirical cross-K function values are \emph{smaller} than the 95\% confidence region for the simulated cross-K values. This implies that (up to at least 250 km) there are \emph{fewer} pairs of points from two groups (BH and FE) compared to what we would expect given independent point patterns. These results suggest a repulsive relationship, where the two spatial point patterns are negatively correlated. Simply put, an act of terrorism by Boko Haram in some location indicates a lower probability of an attack by Fulani extremists in the surrounding area. 

Taken together, the results here suggest that univariate Poisson process models are inadequate for these data, as these would both neglect spatial clustering of attacks \emph{within} each group, and the spatial repulsion of attacks \emph{across} groups. As such, we consider bivariate LGCP models in the following sections, as these accommodate these within and cross-group dynamics.

\subsection{\label{sec:cov_info}LGCP Analysis}

The diagnostics in Section~\ref{sec:detail_estimate_K} suggest a bivariate, inhomogenous LGCP model may be appropriate for the Nigerian terror data. To estimate this model, we first need to supply relevant covariate information to characterize the conditional mean structure of the intensity functions. We focus on plausible determinants of violence that are available at high levels of spatial resolution, including logged population\footnote{Population data is given by Worldpop, which we aggregate to $0.06^{\circ}\times 0.05^{\circ}$.}, elevation\footnote{We use the elevation at the centroid of each grid cell obtained from an \textsf{R} package {\it elevatr}.}, distance to a major city\footnote{For {\it mdis}, we calculate the distance of the grid centroid to the nearest major city (among the 20 most populous cities via {http://www.geonames.org/NG/largest-cities-in-nigeria.html}).}, latitude, longitude, and the interaction between latitude and longitude. These 6 covariates are the inputs to the log-linear function in \eqref{eq:LGCP}, with each standardized to facilitate interpretation.\footnote{Following \cite{gelman2008scaling}, we de-mean each variable and divide them twice by their standard deviation.}

Using these data, we fit univariate and bivariate LGCP models of Nigerian terrorism using our generalization of the {\it lgcp} package in \textsf{R}. As briefly noted above, the current implementation of the LMC structure in {\it lgcp} is overly restrictive for our application, requiring non-negative cross-covariances in bivariate LGCP models. To see this, recall that in \eqref{LMC}, 
\begin{align}\label{eq:cov_ej}
    \mbox{Cov}\{e_i(\mathbf{s}), e_j(\mathbf{u})\}=  \mbox{Cov}\{W_i(\mathbf{s}), W_j(\mathbf{u})\}+\mbox{Cov}\{W(\mathbf{s}), W(\mathbf{u})\},
\end{align} 
for any $i,j=1,2$ and when $i\neq j$, it reduces to
\begin{align}\label{eq:cov_eij}
    \mbox{Cov}\{e_i(\mathbf{s}), e_j(\mathbf{u})\}=  \mbox{Cov}\{W(\mathbf{s}), W(\mathbf{u})\},
\end{align} 
due to the independence of $W_1$ and $W_2$. Thus, $e_1$ and $e_2$ always have non-negative cross-covariance (and cross-correlation), which results in non-negative cross-correlation of $\Lambda_1$ and $\Lambda_2$. 
This is problematic for our application as there is clear separation between the two main terror groups in Figure~\ref{fig:pointpattern}. Therefore, we modify the {\it lgcp} package using a different decomposition of $e_1$ and $e_2$ as 
\begin{align}
e_1= W_1 + W,~~\mbox{and}~~e_2 =  W_2 - W,\label{LMCneg}
\end{align}
where the common process, W, now has a different sign in $e_1$ and $e_2$. From this simple change, it is straightforward to see that with the variance of $e_j$, $\sigma_j^2= \sigma_{W_j}^2+\sigma_W^2$,
\begin{align}
    &\text{Cov}\{e_1(\mathbf{s}),e_2(\mathbf{s})\} =- \text{Var}\{W(\mathbf{s})\}=-\sigma_W^2, \nonumber\\
    &\rho_{12}(h) = -\frac{ \text{Cov}\{W(\mathbf{s}), W(\mathbf{u})\}}{\sqrt{\sigma_{W_1}^2 +\sigma_W^2 }\sqrt{\sigma_{W_2}^2 +\sigma_W^2} }, \label{eq:cor}
\end{align}
and $\rho_{12}$ takes non-positive values for all $h$. As such, researchers can now consider either positive dependence (using \eqref{LMC}) or negative dependence (using \eqref{LMCneg}). As the same parametrization is used in either case, we are able to ensure that first-order intensity functions are free of covariance parameters of $e_1$ and $e_2$ as given in \eqref{constraint}.

With the general LGCP model specified, we can now proceed to estimation using the given data. Initially, we estimate the optimal level of resolution for our analysis, $128\times 128$ square grids over the observational window, using the minimum contrast method (MCM).\footnote{Specifically, preliminary results of $\phi$ using MCM indicate that the grid size should be less than $\hat{\phi}$ (about 12 km). We select $128 \times 128$ as grids satisfying $(2^m) \times (2^n)$ are the most efficient for the Fast Fourier Transform.} Then, using this computational grid and the data, we are able to estimate the model parameters via Bayesian MCMC. Initial values of $\sigma$ and $\phi$ are determined by estimates from MCM (as discussed in Appendix \ref{computation}). Specifically, there are two types of prior densities used in estimation. First, a multivariate Gaussian prior for $\mathbf{\beta}$, with a mean 0 and a large variance (ex. $10^6$) for each component. Second, a multivariate Gaussian prior (on the log scale) for positive $\sigma$ and $\phi$, with the respective means of $\log(\sigma)$ and $\log(\phi)$ given by $\log(1)$ and $\log(\hat{\phi})$ -- where $\hat{\phi}$ is estimated from MCM -- and their variances are both set to 0.15. To estimate the model parameters, we generate $10^6$ iterations as a burn-in period, then we obtain 1000 posterior samples via thinning (every 3000\textsuperscript{th} for univariate analysis and every 5000\textsuperscript{th} for bivariate analysis).

The results for the univariate and bivariate LGCP analysis are reported in Table \ref{tab:fine_uni_biv}, which contains the covariate coefficients, and Table \ref{tab:fine_uni_biv_covariance_par}, which contains the covariance function parameters. For the univariate analysis, we report the results for different samples, first using all terror events (All), before breaking this out by individual terror groups (BH, FE, Other). In the bivariate results, all data is jointly analyzed, with separate marginal results reported for the two main terror groups (BH and FE) and their common process (CP). For each parameter, we report the posterior median (Median) and the 95\% credible interval (Lower and Upper). 
\begin{table}[hbt!]
\begin{center}
\begin{threeparttable}
\caption{Coefficient estimates from LGCP}
\scriptsize
\vspace{-5mm}
\begin{tabular}{llrrrrrr}
\toprule
\multicolumn{2}{c}{} &   \multicolumn{4}{c}{Univariate}
& \multicolumn{2}{c}{Bivariate}\\
\cmidrule(lr){3-6}
\cmidrule(lr){7-8}
Parameter & &All &BH &FE &Other  
&BH &FE \\
\cmidrule(r){1-1}\cmidrule(r){2-6}\cmidrule(r){7-8}
$\text{Intercept}$    
&Lower  
&-9.10  &-12.19    &-10.54    &-11.06    
&-12.21    &-15.89   \\
&Median 
&-8.51  &-11.22    &-9.07    &-10.45    
&-10.99    &-13.88    \\
&Upper 
&-7.76  &-9.94    &-6.39    &-9.76 
&-9.67    &-12.43    \\
\\
$ \log(\text{pop})$
&Lower  
&3.58  &4.50    &1.11    &3.30    
&4.46    &1.09    \\
&Median 
&\textbf{4.10}  &\textbf{5.26}    &\textbf{2.03}    &\textbf{4.08}  &\textbf{5.18}    &\textbf{1.97}\\
&Upper 
&4.64  &6.14    &2.93    &4.92   
&5.85    &2.89   \\
\\
$ \text{elevation}$ 	
&Lower  
&-0.99  &-1.37    &-1.33    &-0.64 
&-0.83    &-1.04   \\
&Median 
&-0.28  &-0.07    &-0.22    &0.16
&0.22    &-0.07    \\
&Upper 
&0.42  &1.02    &0.78    &0.87    
&1.13    &0.93    \\
\\
$\text{mdis}$    
&Lower  
&-2.51 &-3.42    &-3.76    &-1.83 
&-3.72    &-3.23   \\
&Median 
&\textbf{-1.34} &\textbf{-1.70}    &-1.14    &-0.54    
&\textbf{-2.02}    &-0.95   \\
&Upper 
&-0.13  &-0.17    &1.52    &0.69
&-0.58    &1.17   \\
\\
$ \text{lon}$ 	
&Lower  
&2.84  &3.17    &-0.05    &1.02   
&2.89    &-0.33  \\
&Median 
&\textbf{4.50} &\textbf{5.42}    &3.24    &\textbf{2.47}    
&\textbf{4.88}    &3.22\\
&Upper 
&6.03  &7.47    &7.60    &4.25    
&6.96    &7.13   \\
\\
$ \text{lat}$  
&Lower  
&-1.19 &0.67    &-3.07    &-2.32  
&0.39    &-2.52   \\
&Median 
&-0.13 &\textbf{2.20}    &-0.54    &\textbf{-1.29}    
&\textbf{2.34}    &-0.61  \\
&Upper 
&0.97 &4.12    &2.03    &-0.35   
&4.12    &2.06   \\
\\
$ \text{lon*lat}$ 
&Lower  
&-3.38 &-2.45    &-13.27    &-4.72 
&-2.89    &-12.91  \\
&Median 
&-1.02 &1.09    &-6.33    &-1.73
&1.08    &-4.95    \\
&Upper 
&1.54  &4.73    &0.45    &0.83   
&4.43    &1.26    \\
\\
\bottomrule
\end{tabular}
\label{tab:fine_uni_biv}
    \begin{tablenotes}
      \scriptsize
      \item Note: Bold text indicates significant estimates (at 95\% level)
    \end{tablenotes}
\end{threeparttable}
\end{center}
\end{table}

Focusing first on the covariate coefficient results from Table \ref{tab:fine_uni_biv}, we observe several notable results. First, population is the most meaningful input, exhibiting a clear positive association with terror intensity across each of the samples. This is consistent with prior political science research which has found a robust relationship between population and political violence \citep{raleigh2009population}. Second, the results for other covariates are more mixed: elevation is not significantly related to intensity in any of the samples, mdis (i.e., distance to a major urban area) is consistently negatively related to intensity (but not always significantly so), finally the geographic coordinates (lon, lat, and their interaction) vary across samples, reflecting the unique spatial distribution of attacks by different groups.  
\begin{table}[hbt!]
\caption{Covariance parameter estimates from LGCP}
\scriptsize
\begin{center}
\vspace{-5mm}
\begin{tabular}{llrrrrrrr}
\toprule
\multicolumn{2}{c}{} &   \multicolumn{4}{c}{Univariate}
& \multicolumn{3}{c}{Bivariate}\\
\cmidrule(lr){3-6}
\cmidrule(lr){7-9}

Parameter & &All &BH &FE &Other  
&BH &FE & CP \\
\cmidrule(r){1-1}\cmidrule(r){2-6}\cmidrule(r){7-9}
$\sigma$    
&Lower  
&2.08 &2.14    &2.45    &1.57    
&0.99    &1.47    &1.78 \\
&Median 
&2.36  &2.59    &3.10    &1.94   
&1.29    &2.09    &2.28 \\
&Upper 
&2.71  &3.21    &4.11    &2.42    
&1.68    &2.77    &2.67 \\
\\
$\phi$ (Km)  
&Lower  
&28.72  &25.72    &50.72    &19.53 &3.40    &86.26    &57.16 \\
&Median 
&37.55 &38.07    &77.79    &32.37 &6.79    &130.49    &78.43 \\
&Upper 
&52.12 &59.34    &132.15    &51.84 &14.20    &231.75    &111.20 \\
\\
\bottomrule
\end{tabular}
\end{center}
\label{tab:fine_uni_biv_covariance_par}
\end{table} 

While these results are consistent with what could be obtained using conventional Poisson regression models (see Appendix \ref{compare_biv_LGCP_Poisson_regression}), our LGCP analysis provides further information on the process being analyzed. In Table \ref{tab:fine_uni_biv_covariance_par} we report the covariance function parameter results, including $\hat{\sigma}$ and $\hat{\phi}$. Recall that the spatial variance parameter estimate, $\hat{\sigma}^2$, indicates the extent of the variation in $\Lambda(\mathbf{s})$ across the spatial domain (after conditioning on covariates), while the spatial range parameter estimate, $\hat{\phi}$, indicates the size of the spatial clustering.

Focusing first on the spatial variance parameter, $\sigma$, we see that in both the univariate and bivariate analyses the estimates of $\sigma$ for BH are smaller than those for FE. Simply put, this indicates that there is more potential for meaningful spatial clustering in the FE sample. Note that in the bivariate LGCP results, for each sample of events the log of intensity process is decomposed into two parts: the independent marginal process (BH or FE) and the common Gaussian process (CP). Table~\ref{tab:fine_uni_biv_covariance_par} shows that estimate of $\sigma$ for this common process (CP) is larger than each of the respective marginal processes. This indicates the presence of substantial cross-interactions across the two spatial point patterns. 

Next we consider the spatial range paramater, $\phi$, as this also affects the presence and patterns of spatial clusters. Looking first at the univariate results, we see a larger $\hat{\phi}$ value for the FE sample (as compared to that of BH), indicating larger spatial clusters. From the median estimates of $\hat{\phi}$ in Table \ref{tab:fine_uni_biv_covariance_par} we can conclude that \emph{within} each group there is clustering (from $\hat{\phi}$ for BH or FE) up to pairwise distances of roughly $3\hat{\phi}$, where they become independent. For example, in the BH sample the log intensities of pairwise points within $\hat{\phi}=38.07$ have a sample-specific correlation of $\approx 0.37$, indicating positive spatial dependence. However, as the pairwise distance between increases, this correlation decreases. For example, at $3\hat{\phi}$ (or $\approx 114$) the correlation is 0.05, indicating (near) independence. 

For the bivariate analysis, the cluster size of each process is influenced by $\phi$ estimates of marginal (i.e., $\hat{\phi}_{BH}$ or $\hat{\phi}_{FE}$) as well as common process ($\hat{\phi}_{CP}$). Using these we can determine both the \emph{within} group correlation (via \eqref{eq:cov_ej}) and the \emph{across} group correlation (via  \eqref{eq:cor}). Focusing first on the within-group correlation for the BH sample, we see that the log intensities of a pair of points separated by a distance of 6.79 km have marginal correlation $\approx 0.78$, whereas at distance of 200 km the marginal correlation drops to $\approx 0.06$. Similar results obtain for the FE sample, however, these do not approach independence until distances of 250 km and above, implying larger cluster sizes. Finally, we also calculate the bivariate cross correlation by inputting the relevant paramaters from Table \ref{tab:fine_uni_biv_covariance_par} into equation \eqref{eq:cor}. We see that the log intensities have a strongly negative cross-correlation ($-0.64$) at pairwise distance of 0 km, indicating clear dispersion. As before, at greater distances this correlation lessens: at 78.43 km ($\hat{\phi}_{CP}$) the cross correlation is $\approx -0.24$, whereas at three times that distance the cross correlation drops sharply to -0.03.

\begin{figure}[h]
    \begin{subfigure}{0.32\textwidth}
    \includegraphics[width=1\linewidth]{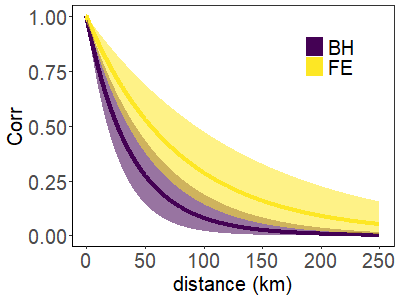}
    \caption{Univariate}
    \label{fig:corr_marginal_uni}
    \end{subfigure}
    \begin{subfigure}{0.32\textwidth}
    \includegraphics[width=1\linewidth]{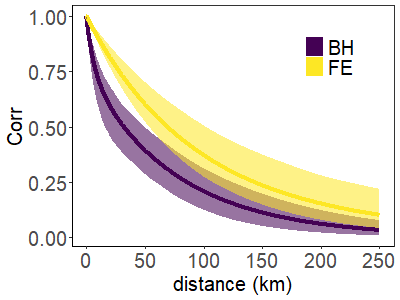}
    \caption{Bivariate}
    \label{fig:corr_marginal_biv}
    \end{subfigure}
    \begin{subfigure}{0.32\textwidth}
    \includegraphics[width=1\linewidth]{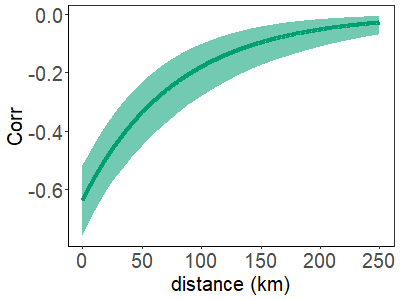}
    \caption{Cross correlation}
    \label{fig:corr_biv_250}
    \end{subfigure}
    \caption{
    Fitted spatial correlation function of log-intensity for univarite and bivariate LGCP models and cross-correlation function of log-intensity for the bivariate models. Shaded regions indicate 95\% credible envelope based on posterior samples.}
\label{fig:corr_marginal} 
\end{figure}

In sum, the bivariate LGCP results indicate positive within-group dependence (clustering) and negative across-group dependence (dispersion), with the magnitude of these effects determined by the radius of the ``neighborhood'' considered. Substantively, this would allow us to determine whether the intensity of attacks in one city correlates with the intensity of attacks in nearby cities.  We can learn more about how these interactions change across distance by plotting the fitted spatial correlation functions against the distance lag, as in Figure \ref{fig:corr_marginal}. In both the univariate (Figure \ref{fig:corr_marginal_uni}) and bivariate models (Figure \ref{fig:corr_marginal_biv}), the spatial correlation of Fulani Extremist attacks is larger than that of Boko Haram attacks, although the difference is not typically significant. The fitted cross-correlation between the two groups from the bivariate analysis (Figure~\ref{fig:corr_biv_250}), on the other hand, is significantly negative. As before, this extends upon what one could learn from a bivariate Poisson regression (see Appendix \ref{compare_biv_LGCP_Poisson_regression} and Table \ref{tab:biv_poisson_Gaussian_copula}), as the cross-correlation parameter estimate -- the comparable quantity to the cross-correlation for spatial lag in Figure~\ref{fig:corr_biv_250} -- is insignificant. 

\subsection{Data limitations and point process models}

There are several unique challenges that arise in the analysis of event data using spatial and spatio-temporal point process methods, including missing events, geolocation error in event reports, low resolution covariate data, repeated events at the same location, etc. Using our illustration, we detail the challenges presented by these issues, and preview how, unlike with standard Poisson regression models, spatial point process models can be generalized to address these issues. Given word limits, we are unable to discuss each of these issues in detail in the main text. Rather than summarize each, we instead discuss one issue -- the resolution of covariate data -- in detail, with the other issues discussed in similar detail in the Appendix. 

\begin{figure}[hbt!]
      \begin{subfigure}{0.32\textwidth} 
      \centering
      \caption{Log population}
    \includegraphics[width=.9\linewidth]{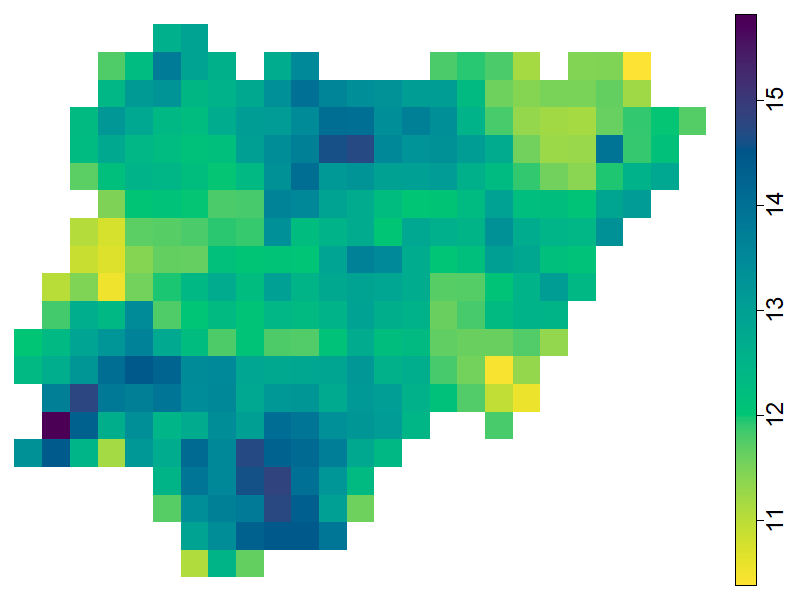}
 \end{subfigure}
    \begin{subfigure}{0.32\textwidth} 
          \centering
    \caption{Elevation}
    \includegraphics[width=.9\linewidth]{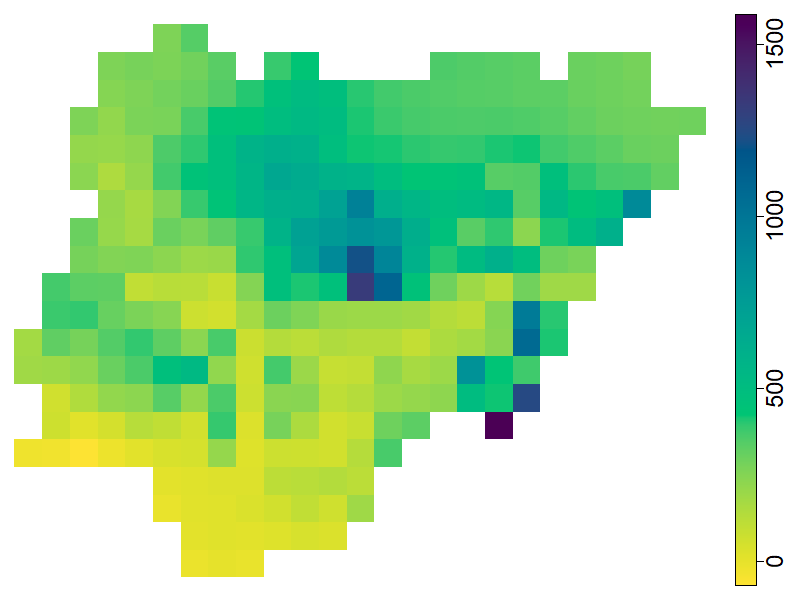}
     \end{subfigure}
          \begin{subfigure}{0.32\textwidth} 
     \centering
          \caption{Distance to city (mdis)}
    \includegraphics[width=.9\linewidth]{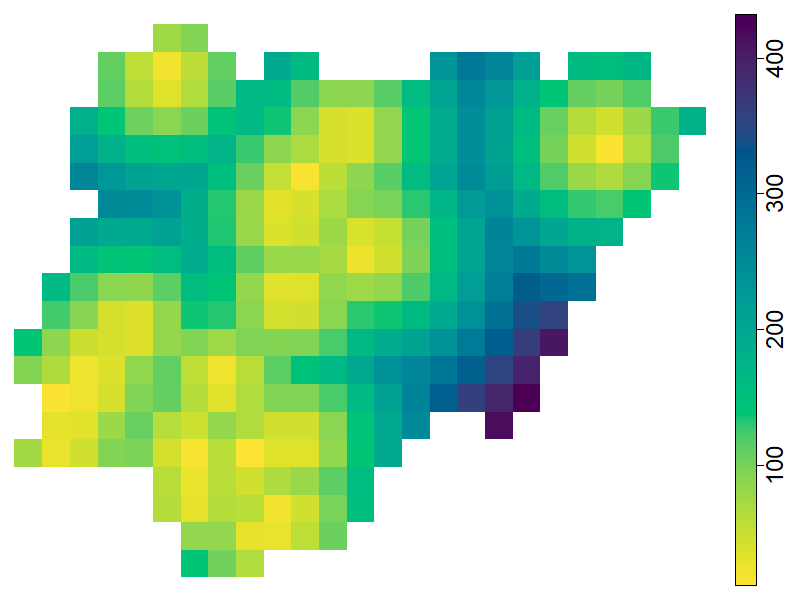}
   \end{subfigure}
         \begin{subfigure}{0.32\textwidth} 
               \centering
    \includegraphics[width=.9\linewidth]{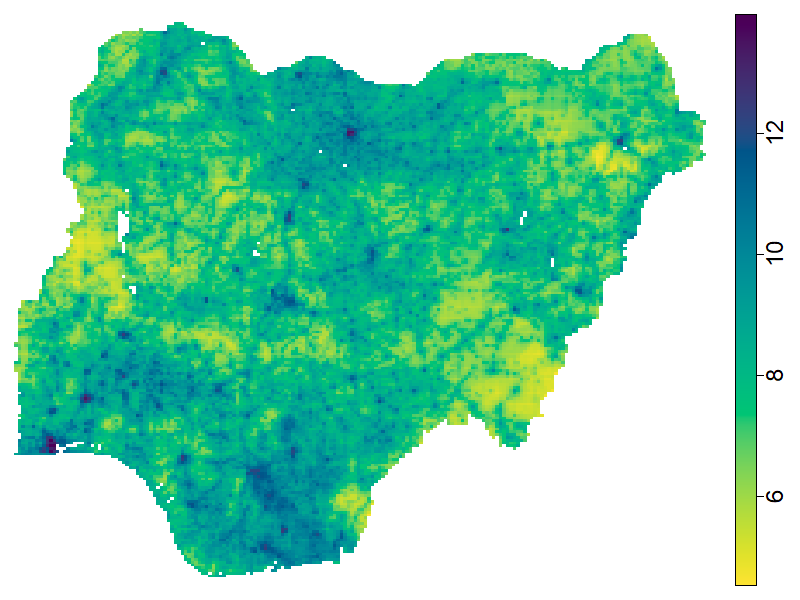}
     \end{subfigure}
    \begin{subfigure}{0.32\textwidth} 
          \centering
    \includegraphics[width=.9\linewidth]{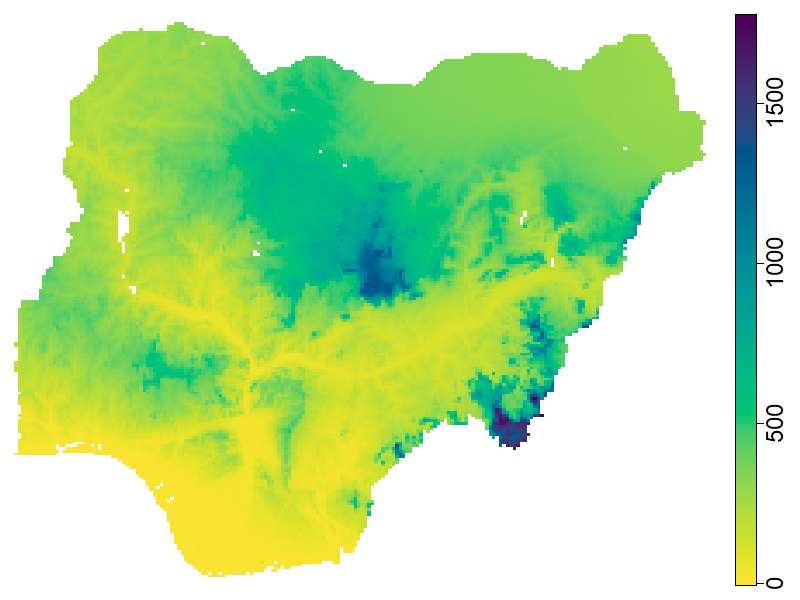}
 \end{subfigure}
      \begin{subfigure}{0.32\textwidth} 
                \centering
    \includegraphics[width=.9\linewidth]{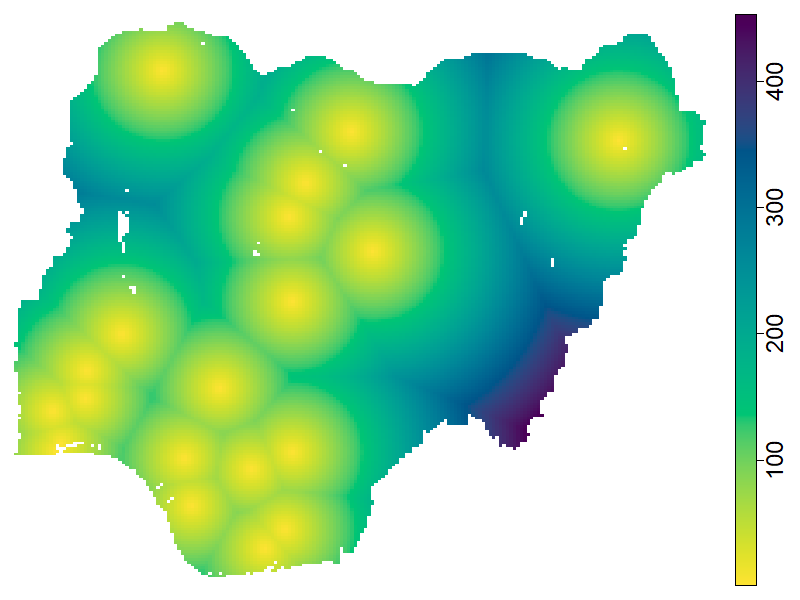}
  \end{subfigure}
   \caption{Covariate data plotted in coarse (top) and fine (bottom) resolution.  }
    \label{fig:coarse_fine_cov_plot_original_scale} 
\end{figure}

Turning back to our Nigerian terrorism application, the main covariates we considered (e.g., population, elevation, and distance to major cities) are available at different spatial resolutions. For other covariates (ex. development), there is often not high-quality, high-resolution data available. In either case, it is important to consider how low resolution inputs might affect spatial point process analyses. In the following analyses, we assess the effect of spatial resolution of covariates by estimating two sets of models: one where we supply covariates at a fine spatial scale (using grids of $0.06 \times 0.05$ degrees, producing 25917 total grid squares), and one where we supply the covariates at a coarser level (using grids of $0.5 \times 0.5$ degrees, producing 312 total grid squares). The difference is the level of resolution can be seen in Figure \ref{fig:coarse_fine_cov_plot_original_scale} where we plot the covariates at the two spatial scales. 

The results from LGCP models supplied with these two sets of covariates are presented in Table~\ref{tab:fine_coarse_mean_biv}. Focusing first on the covariate coefficients, we see that the estimates for each of the significant covariates (ex. log population) is smaller in the coarse data model, indicating the effects may be suppressed due to aggregation at the coarse covariate scale. We also see differences in the covariance function paramater estimates. The radius of clustering is larger in the fine covariate models, as reflected by larger $\hat\phi$. The estimated $\sigma^2_{CP}$ also has a larger proportion (relative to the marginal processes) with fine covariates than that with coarse covariates, indicating that the cross-correlation with fine covariates is larger. Simply put, aggregating the covariate data to the coarse level masks the within-group cluster size and cross-group correlation magnitude in this sample.


\begin{table}[hbt!]
\caption{\label{tab:fine_coarse_mean_biv} Bivariate LGCP estimates with fine vs coarse covariates.} 
\scriptsize
\begin{center}
\begin{threeparttable}
\vspace{-5mm}
\begin{tabular}{llcccccc}
\toprule
\multicolumn{2}{c}{} &   \multicolumn{3}{c}{fine cov (25917 grids)}
& \multicolumn{3}{c}{coarse cov (312 grids)}\\
\cmidrule(lr){3-5}
\cmidrule(lr){6-8}
Parameter & &BH &FE &CP  
&BH &FE &CP \\
\cmidrule(r){1-1}\cmidrule(r){2-5}\cmidrule(r){6-8}
$\sigma$    
&Lower  
&0.99    &1.47    &1.78
&1.96    &2.19    &1.19 \\
&Median 
&1.29    &2.09    &2.28
&2.28    &2.77    &1.53 \\
&Upper 
&1.68    &2.77    &2.67
&2.70    &3.59    &1.94 \\
\\
$\phi$   
&Lower  
&3.40    &86.26    &57.16
&5.57    &61.40    &57.65 \\
&Median 
&6.79    &130.49    &78.43
&8.59    &106.64    &95.36 \\
&Upper 
&14.20    &231.75    &111.20
&12.77    &174.64    &167.30 \\
\cmidrule(r){1-8}
$\text{Intercept}$    
&Lower  
&-12.21    &-15.89   &
&-10.76    &-12.94   & \\
&Median 
&-10.99    &-13.88  & 
&-9.78    &-11.16   & \\
&Upper 
&-9.67    &-12.43   & 
&-8.80    &-9.65    & \\
\\
$ \log(\text{pop})$
&Lower  
&4.46    &1.09   & 
&0.53    &0.41   &\\
&Median 
&\textbf{5.18}    &\textbf{1.97}   &
&\textbf{1.65}    &\textbf{1.90}    &  \\
&Upper 
&5.85    &2.89    &
&2.61    &3.74    & \\
\\
$ \text{elevation}$ 	
&Lower  
&-0.83    &-1.04    &
&-0.58    &-0.65    & \\
&Median 
&0.22    &-0.07   & 
&0.46    &0.19    & \\
&Upper 
&1.13    &0.93    &
&1.43    &1.23    & \\
\\
$\text{mdis}$    
&Lower  
&-3.72    &-3.23    &
&-3.28    &-2.21    & \\
&Median 
&\textbf{-2.02}    &-0.95   & 
&\textbf{-1.81}    &0.20    & \\
&Upper 
&-0.58    &1.17    &
&-0.62    &2.45  &    \\
\\
$ \text{lon}$ 	
&Lower  
&2.89    &-0.33   &
&2.02    &-1.24   & \\
&Median 
&\textbf{4.88}    &3.22   &
&\textbf{3.56}    &2.23   & \\
&Upper 
&6.96    &7.13    &
&5.14    &6.43    &\\
\\
$ \text{lat}$  
&Lower  
&0.39    &-2.52    &
&-0.38    &-2.11   &    \\
&Median 
&\textbf{2.34}    &-0.61 &
&1.35    &0.17    &\\
&Upper 
&4.12    &2.06    & 
&2.86    &2.57    &\\
\\
$ \text{lon*lat}$ 
&Lower  
&-2.89    &-12.91  &
&-0.54    &-11.92  &\\
&Median 
&1.08    &-4.95   &
&2.62    &-4.25   &\\
&Upper 
&4.43    &1.26   & 
&5.86    &2.80   &\\
\\
\bottomrule
\end{tabular}
    \begin{tablenotes}
      \scriptsize
      \item Note: Bold text indicates significant estimates (at 95\% level)
    \end{tablenotes}
\end{threeparttable}
\end{center}
\end{table}

\begin{figure}[hbt!]
\centering
\begin{subfigure}{0.49\textwidth}
    \includegraphics[width=0.9\textwidth]{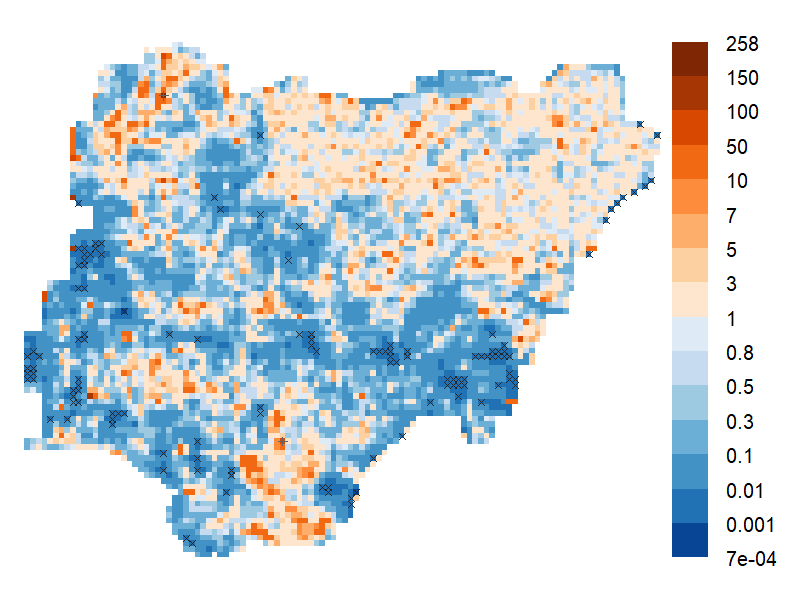}
    \caption{BH}
    \label{fig:biv_bb_median_ratio_contour_cov}
    \end{subfigure}
\begin{subfigure}{0.49\textwidth}
    \includegraphics[width=0.9\textwidth]{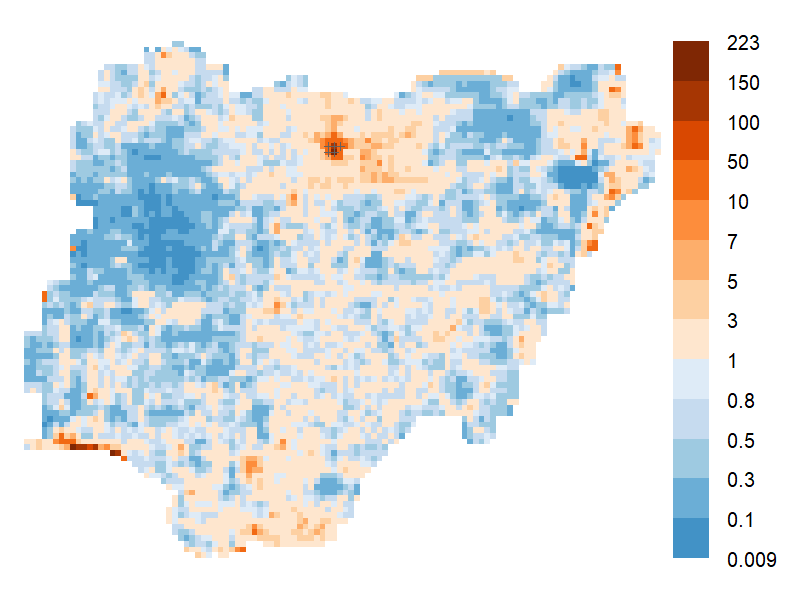}
    \caption{FE}
    \label{fig:biv_ff_median_ratio_contour_cov}
    \end{subfigure}
     \caption{Median ratio of intensity from bivariate LGCP analysis with fine vs coarse covariates. Locations where the ratio is significantly larger than 1 (at 1\% level) are marked $+$, smaller than 1 are marked $\times$. }
    \label{fig:biv-ratio-cov}
\end{figure}

Rather than compare individual coefficients, the overall importance of fine vs. coarse covariates in LGCP modeling can be demonstrated visually. In Figure \ref{fig:biv-ratio-cov} we compare the estimated intensities from bivariate LGCP analysis in the above two cases. Specifically, we simulate ${\Lambda}_j(s)$ 1000 times for both cases (using 1000 sets of posterior estimates $\{\hat{\beta}_{0,j},\hat{\beta}_{1,j},\cdots,\hat{\beta}_{p,j}\}$ and simulated ${e}_j(\mathbf{s})$), calculate the intensity ratio between the two cases in all 1000 simulations, and then plot the median of these 1000 estimated intensity ratios. In Figure \ref{fig:biv-ratio-cov}, values greater than 1 (increasingly red colors) indicate that the estimated intensity was greater in the fine covariate model than the coarse covariate model, values less than 1 (increasingly blue colors) indicate the opposite. This allows us to assess both overall (i.e., global) changes to the intensity under the different sets of covariates \emph{and} location-specific changes as well.

In Figure~\ref{fig:biv-ratio-cov} we see that the increased overall intensity in the fine covariate samples is largely driven by the above \emph{specific} points (the red dots). From this, it seems likely that population -- one of the most significant variables in the analysis -- is the main determinant of these differences, as the intensity ratio is higher in pixels with larger populations. BH has a wider range of the median ratio of intensity across space than FE, in part because the differences of coefficients in two models (fine vs. coarse covariates) for BH are greater than those for FE. Utilising the posterior samples, we also identify locations where the difference between estimated intensities with fine and coarse covariates are significantly larger (marked with $+$) or smaller (marked with $\times$) based on 99\% credible intervals of ratios. For example, in the BH sample the estimated intensity with fine covariates is significantly higher than the estimated intensity with coarse covariates where the population is larger. 
\begin{figure}[hbt!]
\centering
\begin{subfigure}{0.4\textwidth}
    \includegraphics[width=0.9\linewidth]{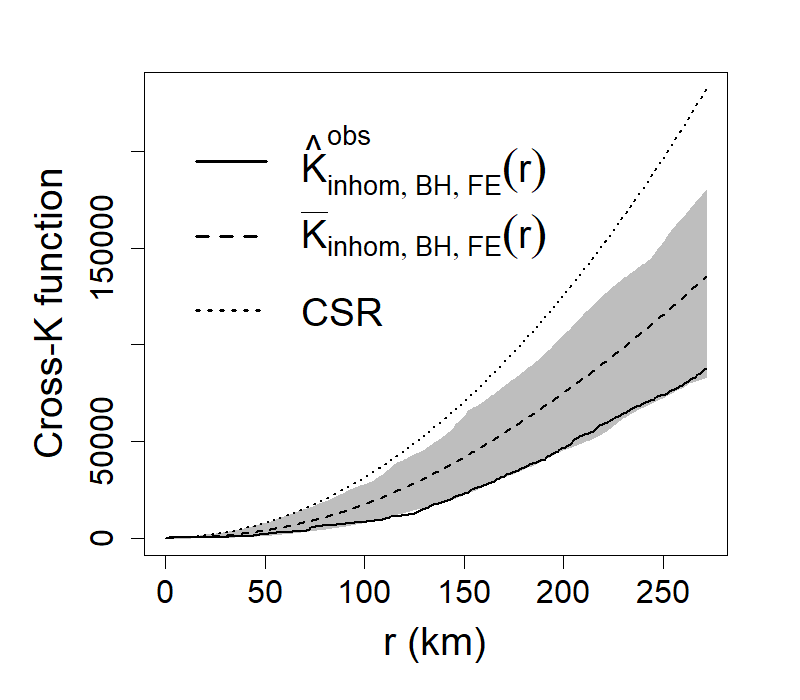}
    \caption{coarse covariates}
    \label{fig:cross_K_coarse}
    \end{subfigure}
\begin{subfigure}{0.4\textwidth}
    \includegraphics[width=0.9\textwidth]{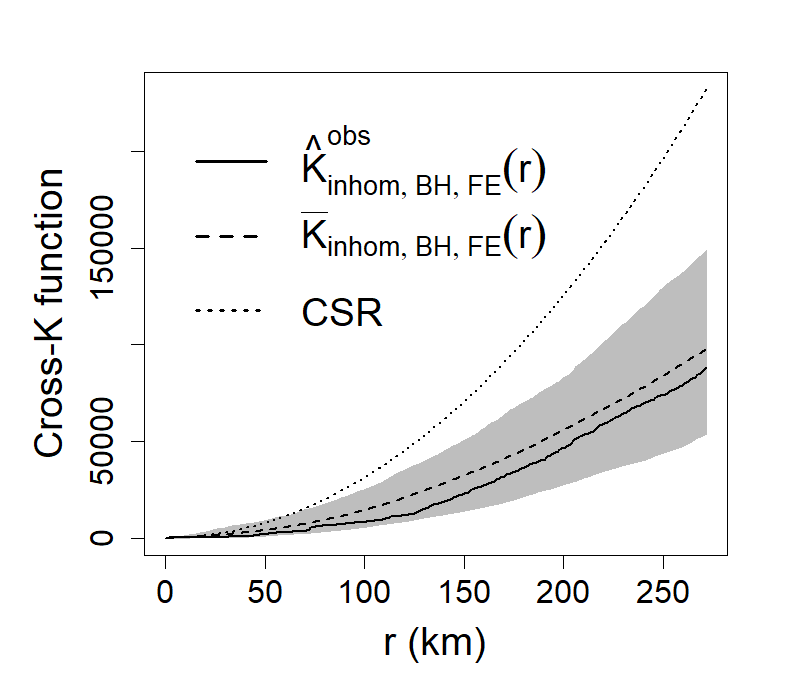}
    \caption{fine covariates}
    \label{fig:cross_K_fine}
    \end{subfigure}
    {\small\caption{ \label{fig:cross_Kinhom}Cross-K functions comparison. Shaded region display 95\% credible regions}}
\end{figure}

The effect of resolution of covariates seems apparent in terms of spatial clustering as well. Figure~\ref{fig:cross_Kinhom} shows the empirical and fitted cross-K function based on bivariate LGCP analysis with coarse and fine covariates. Along with empirical cross-K function, $\hat{K}^{obs}_{\text{inhom}, BH, FE}$, we display  corresponding ``fitted" cross-K functions and their confidence region, based on simulated bivariate LGCP from fitted bivariate LGCP models. For each fitted model, we simulate 180 bivariate spatial point patterns and calculate cross-K function values. Unlike Figure~\ref{fig:crossK_env_mix}, both figures in Figure~\ref{fig:cross_Kinhom} shows better agreement between the empirical curves and the fitted curves, suggesting the importance of the included covariates. Furthermore, the fitted bivariate LGCP with fine covariates show much better agreement between the fitted cross-K function and the empirical cross-K functions. 
Indeed, Figure~\ref{fig:cross_K_fine} demonstrates that the bivariate LGCP model with fine covariates is the most appropriate in describing the interaction between the two spatial point patterns under consideration. 
\begin{figure}[hbt!]
\centering    \includegraphics[width=0.4\textwidth]{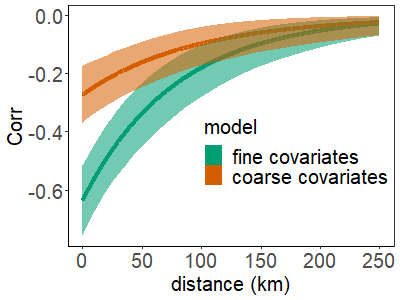}
    \caption{\label{fig:biv_cross_corr}Estimated cross-correlation of log-intensity from bivariate LGCP analysis, with 95\% credible region.}
\end{figure}

Finally, Figure~\ref{fig:biv_cross_corr} shows a comparison between estimated cross-correlation function of log-intensity functions from bivariate LGCP analysis with fine and coarse covariates. For all distances considered, the estimated cross-correlation with coarse covariate is smaller than the estimated cross-correlation with fine covariates. This agrees with findings in Figure~\ref{fig:cross_Kinhom}: cross-K function values of simulated point patterns based on the analysis with coarse covariates are closer to the cross-K function values under CSR, indicating that spatial dependence in the data is masked at this level of aggregation. This demonstrates the importance of high-quality covariate information in both describing the marginal structure for each point pattern \emph{and} accurately reflecting the joint structure in bivariate settings. 

\section{Conclusion}

In this article, we consider the suitability of spatial point process models for political event data. This seems like a natural pairing, as political event data \emph{are} point pattern data. We demonstrate the utility of point process modeling in an analysis of terror attacks in Nigeria. Preliminary diagnostics indicate that the sample(s) exhibit clear spatial dependence. As such, we apply LGCP models to these data, considering both univariate models for individual groups and bivariate analysis for Boko Haram and Fulani Extermists. Ultimately, we find that the bivariate LGCP models best fit the data, demonstrating interesting within group clustering (i.e., positive dependence) and across-group repulsion (i.e., negative dependence). These interesting dynamics are not identified when we analyze all data as a single sample in a univariate model -- i.e., when we fail to differentiate between groups -- as the recovered spatial paramater(s) conflates these within- and between-group effects.  

Despite our advocacy of spatial point process models for political event data, there are a number of limitations in the existing methods that make them ill-suited for these data, and/or limitations in these data which complicate the direct application of existing methods. For example, in existing implementations of the bivariate LGCP model, non-negative cross-correlations are assumed by construction. Since we had strong reasons to suspect negative cross-correlations, we modify the covariance structure utilized in the model. While not a difficult change to implement, this alone complicates the adoption of these methods for applied researchers. Additionally, the data available to researchers is often limited (e.g., covariate scale, geolocation error, identical locations), which can complicate the estimation of point process models or threaten the validity of inferences obtained from such analyses. 

Therefore, broader adoption of point process models in analyzing political event data will either require generalizations to the models or improvements to the data. We have only scratched the surface of that here, and believe the much remains to be done. For example, in future work we aim to generalize LGCP models to explicitly account for geolocation uncertainty in specific events; utilizing event-specific indicators of the geolocation uncertainty that are supplied in event data sets. Aside from spatial accuracy, there are a number of avenues for future development (e.g., more flexible multivariate covariance functions, spatio-temporal extensions, etc.) in this space. We hope that researchers in both political science and statistics will take part in this ongoing dialogue, identifying areas of friction between data and method so that further refinements can be made to spatial point process methods, and deeper insights can be realized from our data.  

\section*{Acknowledgement}

\noindent The authors acknowledge support by NSF DMS-1925119 and DMS-2123247. Mikyoung Jun also acknowledges support by NIH P42ES027704. 

\singlespacing
\bibliographystyle{asa}
\bibliography{reference}

\appendix
\counterwithin{figure}{section}
\counterwithin{table}{section}
\counterwithin{equation}{section}
\counterwithin{footnote}{section}

\begin{appendices}

	\newpage
	\section{\label{LGCP_cov_sims} LGCP spatial paramaters and patterns}
	
	To better demonstrate the effects of $\phi$ and $\sigma$ on spatial point patterns from LGCP models, we simulate four spatial point patterns from LGCP models under different values of $\phi$ (0.1 vs. 4) and $\sigma$ (0.1 vs. 1) over the spatial domain of $[0,10] \times [0,10]$. We simulated LGCP processes using {\it simLGCP} function in an \textsf{R} package {\it geostatsp} \citep{geostatspPackage}. Across all simulations, we set $\beta_i=0$ for all $i$'s in \eqref{eq:LGCP} and use an exponential correlation function for the covariance structure of $e$. Recall that under this setting $\tilde{\lambda}=1$, thus the expected number of simulated points in each sample is 100.

	\begin{figure}[hbt!]
		\includegraphics[width=\linewidth]{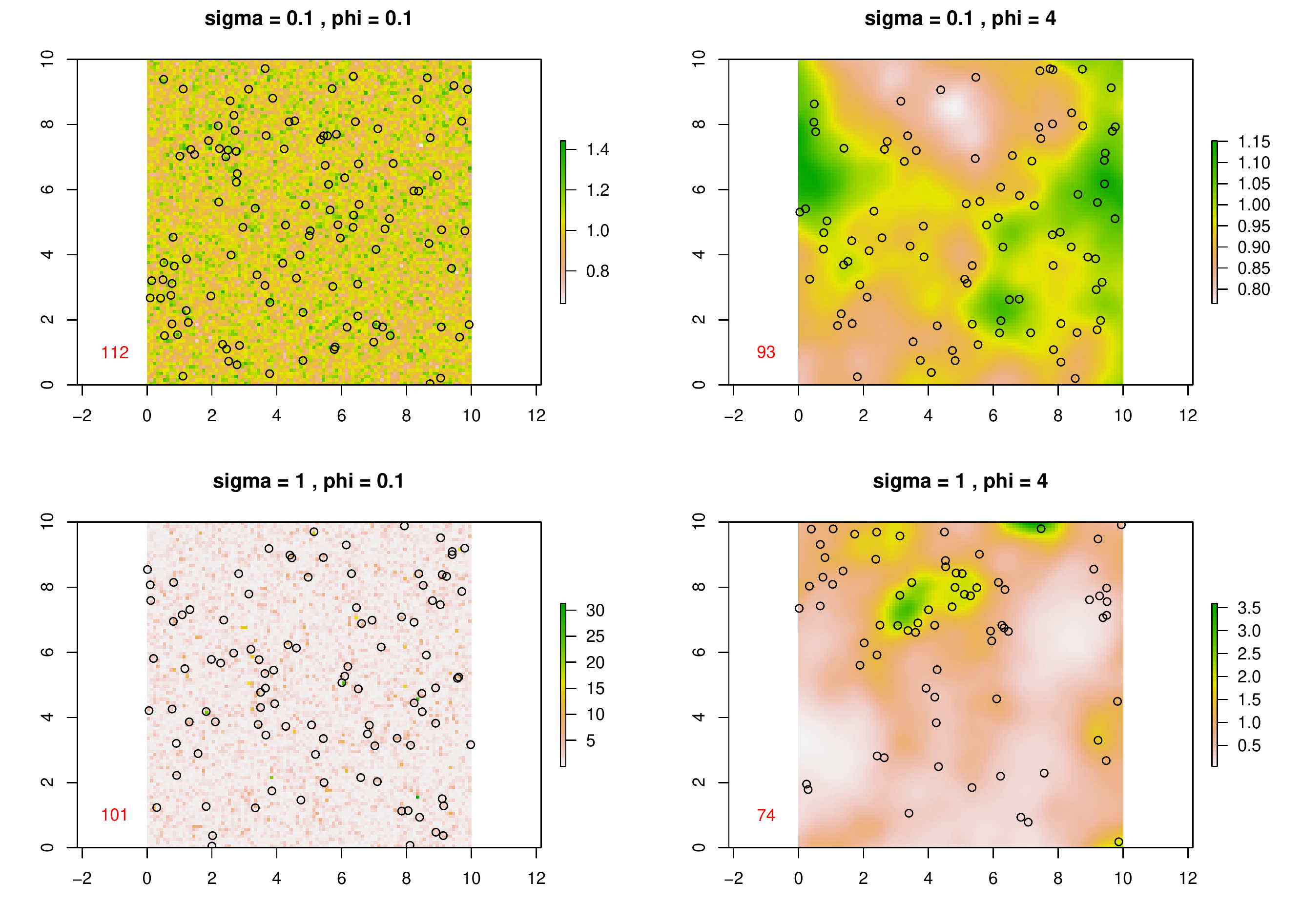}
		\caption{Four simulated spatial point patterns from LGCP models. For each panel, a simulated intensity field, $\Lambda$, is displayed in the background. The number in red indicates the number of simulated points. }
		\label{fig2-test}
	\end{figure}
	
	The results from the simulations are given in Figure~\ref{fig2-test}, with the different panels mapping the sample of points produced under different values of $\sigma$ (varying across the grid rows) and $\phi$ (varying across the grid columns).\footnote{
		Note that for all 4 panels, numbers of simulated points are around 100, as expected.
	} As noted in the main text, when $\sigma$ is small (top row), the $e$ term has relatively small influence over $\Lambda$ and thus $\Lambda$ does not vary much over space. Conversely, when $\sigma$ is large (bottom row), $\Lambda$ is heavily influenced by the variation in $e$, and thus simulated $\Lambda$ shows clear spatial variation across the domain. This effect can now be clearly seen in Figure~\ref{fig2-test}. When $\sigma$ is low (top row), the simulated $\Lambda$ values do not vary much across space and their values are around $\tilde{\lambda}=1$. Therefore, increasing $\phi$ from 0.1 (top left) or 4 (top right) has little consequence, since the low $\sigma$ constrains the influence that even a large $\phi$ could have on the clustering in the simulated point pattern. 
	
	When $\sigma$ is large (bottom row), however, $\Lambda$ is substantially affected by the variation in the $e$ term. As such, we would expect that the simulated $\Lambda$ values vary widely across the spatial domain. Under these conditions, variation in $\phi$ has a significant effect on the cluster patterns we observe. When $\phi$ is small (bottom left), we do not observe spatial clustering between points, even with a large $\sigma$. However, when $\phi$ is large (bottom right), we see large and meaningful spatial clusters. 
	Tying this back to the models discussed above, the bottom-right panel is an example where the Poisson process model would be clearly insufficient and an LGCP model should instead be preferred.

	\newpage 
	\section{\label{computation} Fitting point process models}
	\setcounter{figure}{0}
	
	
	The covariance parameters of log-intensity functions in LGCP models (i.e., $\mathbf{\beta}$, $\sigma^2$ and $\phi$) can be estimated in a few different ways. One is a moment-based method, introduced in \cite{moller_et_al98}, called {\it Minimum Contrast Method} (MCM) that utilisizes the Ripley's K function. That is, it measures the discrepancy between empirical and theoretical second-order quantities such as inhomogeneous K function in \eqref{eq:Kinhom(r)} and obtain the estimates of parameters as the minimizer of this discrepancy. 
	
	\begin{figure}[h]
		\centering
		\includegraphics[width=0.4\textwidth]{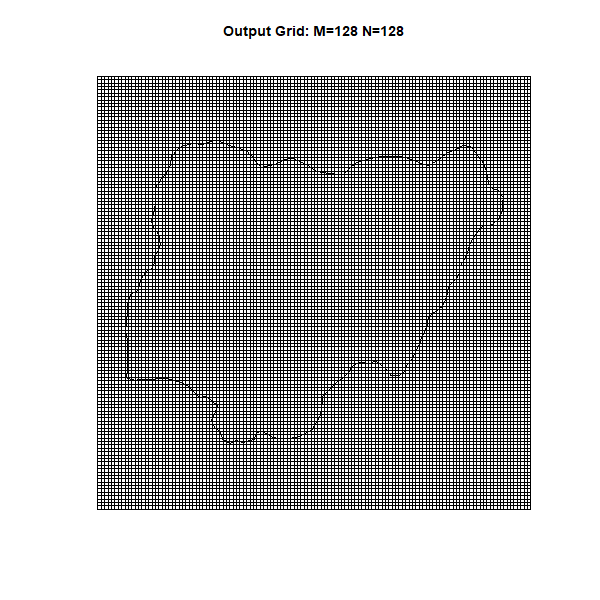}
		\caption{Observational window with computational grid (cell width 12 km) used for LGCP computation}
		\label{fig:obs_window}
	\end{figure}

	More statistically efficient ways utilize likelihood functions of LGCP. Likelihood calculation of LGCP models is challenging due to the doubly stochastic nature of the process. Specifically, the intensity function of a Poisson process is also stochastic. As discussed in \cite{diggle2013spatial}, for a point pattern $\mathbf{x}=\{\mathbf{x}_i \in  D:i=1,\ldots,n\}$ generated from LGCP models, the likelihood is written as
	\begin{align}
		L(\mathbf{\theta};\mathbf{x})&=P(\mathbf{x}|\mathbf{\theta})=\int_{\Lambda}P(\mathbf{x},\Lambda|\mathbf{\theta})\text{d}\Lambda = \mathbb{E}_{\Lambda|\mathbf{\theta}}\{L^{*}(\Lambda;\mathbf{x})\} \quad \text{with} \label{eq:likelihood}\\
		l^{*}(\Lambda;\mathbf{x})&=\log L^{*}(\Lambda;\mathbf{x})=\sum^n_{i=1}\Lambda(\mathbf{x}_i)-\int_D\Lambda(s)\text{ds}, \label{eq:likelihood1}
	\end{align}
	where $\mathbf{\theta}=(\mathbf{\beta},\sigma^2,\phi)$. It involves integrating the stochastic likelihood $P(\mathbf{x},\Lambda|\mathbf{\theta})$ (because of stochastic intensity process $\Lambda$) over the infinite-dimensional distribution of $\Lambda$, which is challenging. One can approximate the integral in \eqref{eq:likelihood} with Monte Carlo approximation of the likelihood \citep{jun_et_al19}. 
	
	In this paper, we use Bayesian inference for the LGCP models. We use an \textsf{R} package, {\it lgcp}, that provides tools for univariate and bivariate LGCP analysis with Bayesian MCMC \citep{taylor2015bayesian}. Likelihood calculation of LGCP models either for Monte Carlo approximation or Bayesian MCMC require numerical approximation of integral in \eqref{eq:likelihood1}. Instead of integrating the intensity function over the domain D analytically, we place a fine regular grid over the domain and approximate the integral by a Riemann sum. Figure~\ref{fig:obs_window} displays the computational grid ($128 \times 128$) determined by internal function in the {\it lgcp} package.
	
	\newpage
	
	\section{\label{compare_biv_LGCP_Poisson_regression} Bivariate Poisson regression with copula}
	
	To see how bivariate LGCP models compare with copula-based bivariate count regression, we sum the number of events in each spatial grid for two main groups (BH, FE) and use a copula-based bivariate Poisson regression model.\footnote{For more details about copula regression, please see \citep{cameron2004modelling,trivedi2007copula}} A Gaussian copula function is used to model the dependence of the bivariate count data.\footnote{An \textsf{R} package {\it GJRM} is used for the analysis \citep{GJRM-R-package-web}} PRIO spatial grid resolution, $0.5^{\circ} \times 0.5^{\circ}$, with total 312 grid cells are used. In Table \ref{tab:biv_poisson_Gaussian_copula}, two columns represent coefficients in each marginal Poisson regression of counts for BH and FE respectively, and $\theta$ is the cross correlation (dependence) parameter. 
	
	\begin{table}[hbt!]
		\caption{\label{tab:biv_poisson_Gaussian_copula} Parameter estimates (with their standard errors) for bivariate Poisson regression with counts data. Bold faced indicate significant values (at 5\% level).}
		\centering
		\begin{tabular}{l|rr}
			\hline
			Parameter &  BH & FE \\ \hline
			$\text{Intercept}$ 
			& \textbf{$-$1.50 (0.13)}
			& \textbf{$-$1.18 (0.12)}
			\\
			$\text{log(pop)}$ 
			& \textbf{0.60 (0.12)}
			& \textbf{$-$0.54 (0.26)}
			\\
			$\text{elevation}$ 
			& 0.18 (0.14)
			& \textbf{0.31 (0.15)}
			\\
			$\text{mdis}$ 
			& \textbf{$-$1.87 (0.16)}
			& \textbf{$-$1.92 (0.32)}
			\\
			$\text{lon}$ 
			& \textbf{4.94 (0.25)}
			& \textbf{2.22 (0.41)}
			\\
			$\text{lat}$ 
			& \textbf{0.66 (0.26)}
			& \textbf{$-$1.52 (0.23)}
			\\
			$\text{lon*lat}$ 
			& \textbf{$-$2.62 (0.57)}
			& \textbf{$-$6.00 (0.84)}
			\\
			\hline
			\multicolumn{1}{c|}{$\theta$}
			& \multicolumn{2}{c}{$-0.05$ (0.07)}\\
			\hline
		\end{tabular} 
	\end{table}
	
	\begin{figure}[H]
		\centering
		\begin{subfigure}{0.4\textwidth}
			\includegraphics[width=1\textwidth]{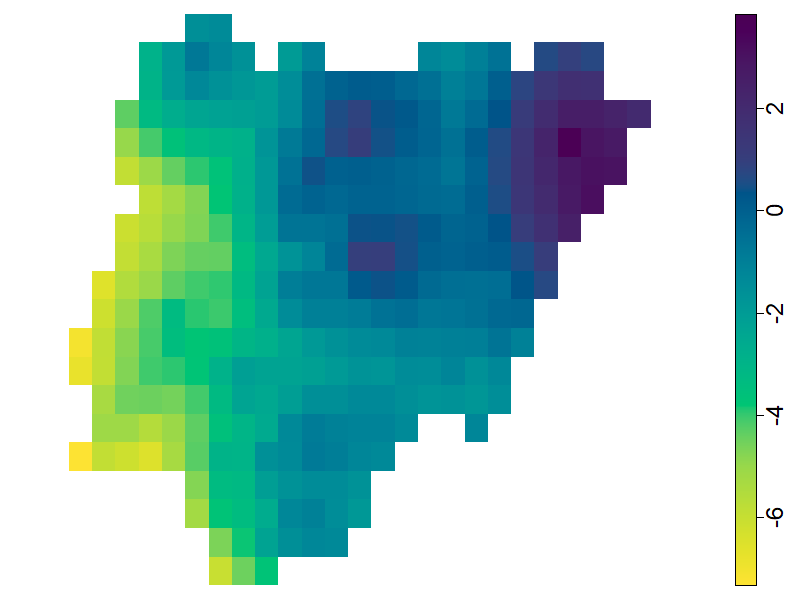}
			\caption{BH}
			\label{fig:bb_linear_trend_copula}
		\end{subfigure}
		\begin{subfigure}{0.4\textwidth}
			\includegraphics[width=1\textwidth]{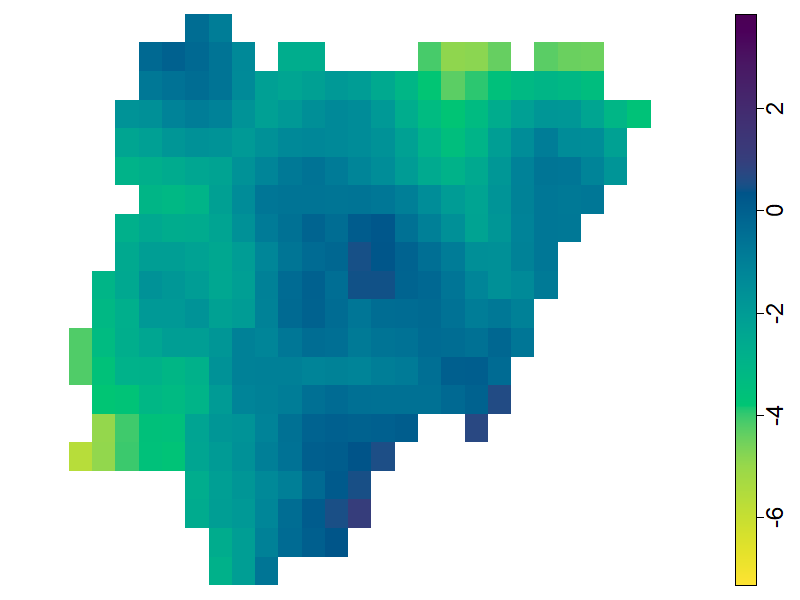}
			\caption{FE}
			\label{fig:ff_linear_trend_copula}
		\end{subfigure}
		
		\caption{Log of mean count from bivariate Poisson with copula.}
		\label{fig:linear_pred_biv_poisson_copula}
	\end{figure}
	
	Differing from the results of bivariate LGCP (Table \ref{tab:fine_coarse_mean_biv}), Table \ref{tab:biv_poisson_Gaussian_copula} shows bivariate Poisson regression with copula has significantly negative coefficient of \texttt{log(pop)} for FE. This may because FE has many zero counts where \texttt{log(pop)} are relatively large (such as the northern and northeastern regions of Nigeria). Compared to the LGCP results, bivariate Poisson with copula also obtains a negative cross correlation parameter $\theta$, although it is not significantly different from zero. Also note that the bivariate LGCP and bivariate Poisson with copula both obtain similar results for \texttt{lon},  \texttt{lat}, and their interaction, reflecting the similar overall spatial patterns. The patterns from the bivariate Poisson are given in Figure \ref{fig:linear_pred_biv_poisson_copula}, which reflect that BH has higher density in the northeastern while FE has higher density in the center of Nigeria.

	\clearpage 
	\newpage 
	\section{Bivariate Robustness}
	
	In the main text, we briefly presented several limitations in social events data which make them ill-suited for spatial point process analysis: resolution of covariates, identical locations, and geolocation specificity. In the main text, we focused on the spatial resolution of the covariates in some detail, using plots of the cross-K function and cross-correlation envelope of logged intensity process in bivariate LGCP models to conclude that using finer resolution of covariates is very important in point process analysis. Here we similarly elaborate on two additional issues encountered when analyzing event data with point process methods: common locations and geolocation error. 
	
	\subsection{Events with common locations}
	
	In political event data there are often events recorded with \emph{identical} spatial location information. For the GTD data used in our illustration, there are multiple events with the exact same longitude and latitude information\footnote{Note that this is distinct from duplicate event reports, which are often pre-processed and removed -- i.e., de-duplicated -- before the event data are initially released.} This occurs with some regularity in political events data since the raw information (e.g., news reports) often contain imprecise information on the location of each event. In the absence of precise coordinate information, events are commonly coded as occurring in the centroid of the spatial location (e.g., city, district) identified in the original source. This can produce two problems for analysts: 1) points with identical locations and not permitted in point process models, 2) measurement error in locations can produce biased estimators of spatial covariance and covariate parameters of interest. We take up the first issue here (identical locations) and save the latter (measurement error) for the next subsection. 
	
	\begin{table}[htb!]
		\caption{Number of terrorism attacks for Nigeria in 2014}
		\centering
		\vspace{-3mm}
		\begin{tabular}{l|r|r}
			\hline
			& Total number & Attacks with identical location \\
			\hline
			BH  & 436 & 152 \\
			FE & 156 & 15 \\
			Other & 122 & 16 \\
			\hline 
			All & 714 & 190 \\
			\hline
		\end{tabular}
		\label{tab:num_total_drop}
	\end{table}
	
	\par To get a sense of the severity of this problem, in Table \ref{tab:num_total_drop} we report the number of events with identical locations in the GTD data for Nigeria in 2014. Across all groups, we observe that 190 of the 714 have another event reported with exact same spatial coordinates. For attacks by Boko Haram, in particular, 152 of the 436 events share a location with another event.\footnote{Note that there are 6 events where Boko Haram and Unknown share a common location, and 1 event where Fulani extremists and Unknown do. This is why the group-specific totals reported (152, 15, and 16) do not sum to the All groups total (190).} Depending on the level of aggregation, this may not be a large problem when analyzing areal data using count data models, since precise coordinate information is not required. However, when using spatial point pattern analysis data with multiple identical locations is quite problematic, as the probability of having two events at the exact same location is zero. As such, samples of points with unique locations are typically required for point process methods. 
	
	To satisfy this requirement with political event data, researchers are forced to make ad hoc decisions in data processing prior to the analysis. One approach is to eliminate repeated location events from the analysis. While this ensures unique coordinates for the remaining events, it risks sample selection bias. For example, for all attacks from the Nigeria data reported in Table \ref{tab:num_total_drop} this would eliminate 190 events from the sample. Another approach is to  add small random numbers to the coordinate information to jitter spatial locations \citep[e.g.,][]{bridwell2007dimensions, tompson2015uk}. While this strategy retains the entire sample of events, it also induces additional measurement error into this retained subset of events.   
	
	To see the consequences of these approaches, we re-estimate our bivariate LGCP models with jittered vs non-jittered samples. To construct the jittered sample, we add noise to the longitude and latitude for the observations; taking draws from a normal distribution with mean zero and standard deviation $10^{-6}$ (in degrees), and adding these to the original longitude and latitude. We then fit bivariate LGCP models for two cases: (1) all data with jittered locations (2) data with only unique locations (i.e., omitting repeated location events). Estimates of the parameters for both cases are reported in Table~\ref{tab:fine_biv_consider_identical_locations}.

	While the covariate coefficient results are quite similar in both cases, the covariance parameters of log-intensity functions vary quite a bit. Specifically, considering jittered sample has larger $\hat{\sigma}$ and smaller $\hat{\phi}$, we find that for jittered sample than those for non-jittered sample, the within-group cluster size is smaller but the variation of realised $\Lambda_j(\mathbf{s})$ ($j=1,2$) across its expectation $\tilde{\lambda}_j(\mathbf{s})$ is larger, and the cross-correlation is larger in small spatial lags (e.g. $\leq 90$ km) but is smaller when distance increases. The reason the non-jittered sample indicates a larger cluster size and smaller magnitude of spatial cross-correlation in close distances is because of the omission of a subset of cases which happened in close proximity to some remaining cases. Even if we have reason to doubt that the repeated events occurred at the \emph{same} location, we know from the reports that they occurred within reasonably close proximity to one another. As such, removing these from the sample will increase the within-group cluster size and decrease the cross-group correlation in close distances by construction and lead to a more flat intensity surface. It should be avoided, especially where researchers care theoretically about modeling spatial dependence or the global changes of the intensity surface. 
	
\begin{figure}[hbt!]
		\centering
		\begin{subfigure}{0.49\textwidth}
			\includegraphics[width=0.9\textwidth]{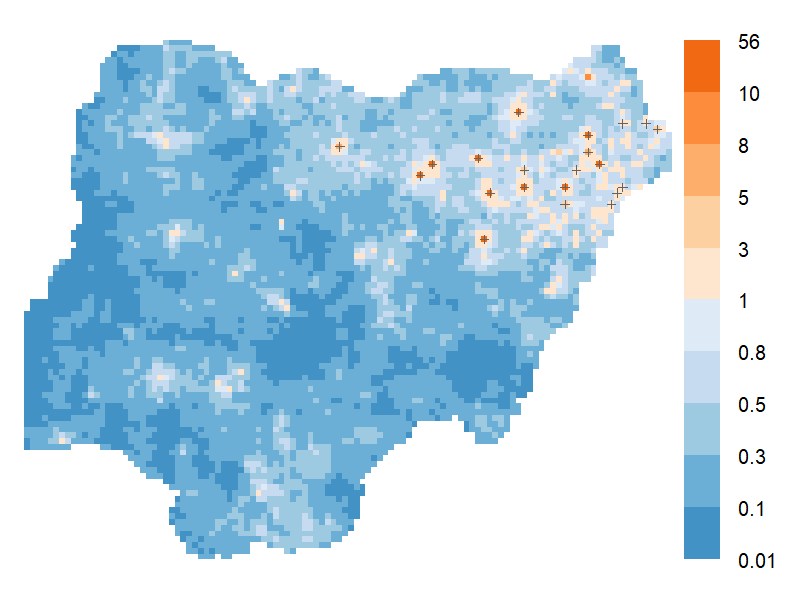}
			\caption{BH}
			\label{fig:biv_bb_median_ratio_contour_jitter}
		\end{subfigure}
		\begin{subfigure}{0.49\textwidth}
			\includegraphics[width=0.9\textwidth]{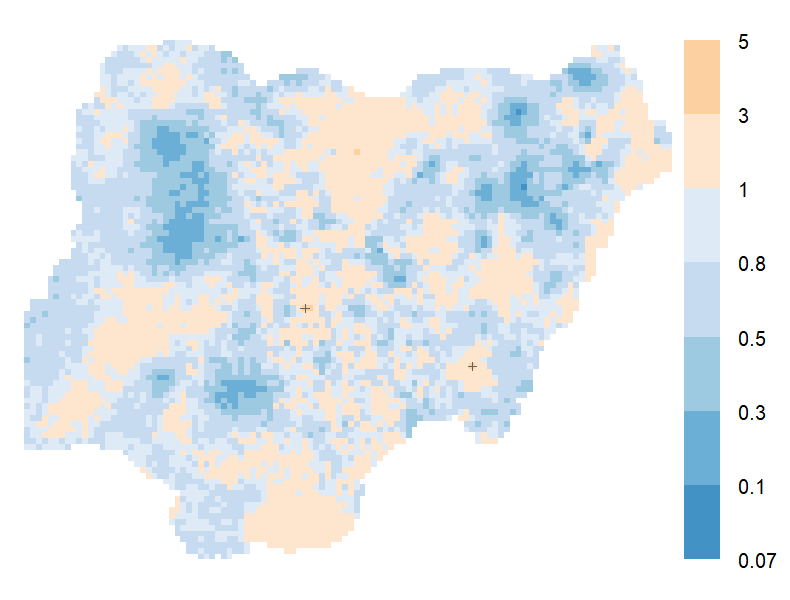}
			\caption{FE}
			\label{fig:biv_ff_median_ratio_contour_jitter}
		\end{subfigure}
		\caption{Median ratio of intensity from bivariate LGCP analysis with jittered vs unique locations. Locations where the ratio is significantly larger than 1 (at 1\% level) is marked with $+$ and those with ratio significantly smaller than 1 are marked with $\times$. }
		\label{fig:ratio-jit-biv}
	\end{figure}

	The consequence of these choices also affects overall model predictions, as shown in Figure~\ref{fig:ratio-jit-biv}. As in Figure~\ref{fig:biv-ratio-cov} above, here we plot the ratio (jittered over non-jittered) of fitted log intensities to compare the two models. Looking first to the Boko Haram results in panel (a), the median intensity ratio across space has a range $[0.01, 56]$. The median and mean of these values are 0.23 and 0.34 respectively. We see two patterns: 1) in areas with many attacks (i.e., the northeastern part of the country), the estimated intensity is substantially larger for jittered sample, and 2) in areas with fewer attacks, the estimated intensity is larger for the non-jittered sample (the blue areas). This result reflects the fact that in the non-jittered sample the estimated intensity is more uniform across space given the omission of cases in high intensity areas. With the jittered sample, these cases are retained and we are able to better differentiate high- and low-intensity areas. This is similar for the Fulani Extremists events, the median intensity ratio across space has a range\footnote{This range is smaller than that for Boko Haram because Fulani Extremists events only have 15 points which will be removed in non-jittered sample.} $[0.07, 5]$. The median and mean of these values are 0.82 and 0.86 respectively. We see that the estimated intensity is larger for jittered cases in the center of the country where these attacks commonly occur. However, for both groups, the ratio of intensities are not significantly different from 1 (at the significance level of 1\%), except for a few pixels in the eastern part of country in the Boko Haram models.

	\begin{table}[hbt!]
		\caption{\label{tab:fine_biv_consider_identical_locations} Bivariate LGCP estimates with jittered vs non-jittered locations.} 
		\scriptsize
		\begin{center}
			\begin{threeparttable}
				\vspace{-5mm}
				\begin{tabular}{llcccccc}
					\toprule
					\multicolumn{2}{c}{} &   \multicolumn{3}{c}{Jittered}
					& \multicolumn{3}{c}{Non-jittered}\\
					\cmidrule(lr){3-5}
					\cmidrule(lr){6-8}
					Parameter & &BH &FE &CP 
					&BH &FE &CP\\
					\cmidrule(r){1-1}\cmidrule(r){2-5}\cmidrule{6-8}
					$\sigma$    
					&Lower  
					&0.99    &1.47    &1.78 
					&0.64    &1.84    &1.31 \\
					&Median 
					&1.29    &2.09    &2.28 
					&0.96    &2.41    &1.67 \\
					&Upper 
					&1.68    &2.77    &2.67  
					&1.24    &3.19    &2.05  \\
					\\
					$\phi$   
					&Lower  
					&3.40    &86.26    &57.16  
					&4.21    &71.65    &66.17   \\
					&Median 
					&6.79    &130.49    &78.43  
					&8.53   &115.64    &101.26  \\
					&Upper 
					&14.20    &231.75    &111.20  
					&15.28    &195.48    &161.97   \\
					\cmidrule{1-8}
					$\text{Intercept}$    
					&Lower  
					&-12.21    &-15.89   &
					&-11.88    &-13.63   &\\
					&Median 
					&-10.99    &-13.88   &
					&-10.98    &-11.93   &\\
					&Upper 
					&-9.67    &-12.43   &
					&-10.13    &-10.48  & \\
					\\
					$ \log(\text{pop})$
					&Lower  
					&4.46    &1.09   &
					&3.40    &1.06   & \\
					&Median 
					&\textbf{5.18}    &\textbf{1.97}  &
					&\textbf{3.96}    &\textbf{1.96}  &\\
					&Upper 
					&5.85    &2.89   &
					&4.63    &2.83   & \\
					\\
					$ \text{elevation}$ 	
					&Lower  
					&-0.83    &-1.04   &
					&-0.79    &-0.81   &    \\
					&Median 
					&0.22    &-0.07   &
					&0.16    &0.06     &\\
					&Upper 
					&1.13    &0.93   &
					&1.03    &0.92   & \\
					\\
					$\text{mdis}$    
					&Lower  
					&-3.72    &-3.23   &
					&-3.36    &-3.11   &    \\
					&Median 
					&\textbf{-2.02}    &-0.95  & 
					&\textbf{-1.91}    &-0.74  &  \\
					&Upper 
					&-0.58    &1.17  & 
					&-0.55    &1.38  & \\
					\\
					$ \text{lon}$ 	
					&Lower  
					&2.89    &-0.33   &
					&2.07    &-0.83   & \\
					&Median 
					&\textbf{4.88}    &3.22  &
					&\textbf{3.96}    &2.75  &\\
					&Upper 
					&6.96    &7.13   &
					&5.68    &6.72   & \\
					\\
					$ \text{lat}$  
					&Lower  
					&0.39    &-2.52   &
					&0.24    &-2.66   & \\
					&Median 
					&\textbf{2.34}    &-0.61  &
					&\textbf{1.73}    &-0.23  &\\
					&Upper 
					&4.12    &2.06   & 
					&3.41    &2.35   & \\
					\\
					$ \text{lon*lat}$ 
					&Lower  
					&-2.89    &-12.91   &   
					&-2.20    &-12.31   &  \\
					&Median 
					&1.08    &-4.95   & 
					&0.81    &-4.38   & \\
					&Upper 
					&4.43    &1.26   & 
					&3.83    &1.85   & \\
					\\
					\bottomrule
				\end{tabular}
				\begin{tablenotes}
					\scriptsize
					\item Note: Bold text indicates significant parameter estimates (at 95\% level)
				\end{tablenotes}
			\end{threeparttable}
		\end{center}
	\end{table}

	\subsection{Geolocation error in event data}
	
	Next, a related but different issue common in political events data is imprecise spatial coordinate information \citep{weidmann2016closer, lee_liu_ward_2019}. Precisely evaluating the extent of this problem is difficult without ground-truth data (as in \cite{weidmann2015accuracy}). However, many event datasets provide indicators on the accuracy of the supplied spatial coordinates for each observation that we can use these to get a rough approximation of the magnitude of this issue. The GTD data, for example, includes a variable {\it specificity} which indicates how precise the geolocation information is for each event. For instance, $\text{specifity} = 1$ represents the event occurred in city/village/town, and the exact coordinate information (longitude and latitude) is given. Specificity values larger than 1 indicate less certainty in the geolocation information, up to `5' where ``coordinate info unknown.'' 

	Table \ref{tab:spec} describes how specificity variable is defined in GTD and numbers of events by specificity and by group in our Nigeria 2014 sample. Both attacks by BH and FE have a non-negligible proportion of imprecise locations. In the following, we demonstrate the effect of such geolocation uncertainty by comparing LGCP results of events with specificity $=1$ vs. specificity $>1$. Uncertainty in geolocation is expected to affect the results of point process analysis -- particularly for spatial clustering -- since these coordinates \emph{are} the random variable of interest. Moreover, because exact spatial coordinates are still supplied for cases with low specificity, we can observe cases for which we have little information (ex. specificity = 4) being recorded as if they occurred at identical locations (as discussed above). As a result, the inter-distance between two events in the case of specificity $>1$ will often be smaller than the other case. Indeed, the mean distance between pairs of events for BH and FE are 274.9 km and 244.0 km, respectively, for specificity$=1$, while for specificity $>1$, the corresponding values are 192.1 km and 236.9 km.

	\begin{table}[hbt!]
		\caption{Description of specificity and number of attacks according to its specificity}
		\label{tab:spec}
		\centering
		\begin{tabular}{c|c||rrr}
			\hline
			Specificity&Description&BH&FE&Other\\
			\hline 
			\hline
			1& Exact location known &302&70&90\\
			\hline
			\multirow{3}{*}{2} & Occurred in city/village/town, 
			&\multirow{3}{*}{98} &\multirow{3}{*}{62}
			& \multirow{3}{*}{14} 
			\\
			& coordinates  of center of the & & & \\
			& smallest administrative region given & & & \\
			\hline
			\multirow{3}{*}{3} & Info on city/village/town unknown,  
			&\multirow{3}{*}{18} &\multirow{3}{*}{21}
			& \multirow{3}{*}{7} \\
			& coordinates of center of the & & &  \\
			& smallest administrative region given & & &  \\
			\hline
			\multirow{3}{*}{4} & Info on 2nd order or smaller region 
			&\multirow{3}{*}{18} &\multirow{3}{*}{3}
			& \multirow{3}{*}{11} 
			\\
			& unknown, coordinates of the center of  & & & \\
			& 1st order administrative region given & & & \\
			\hline
			5& coordinate info unknown& 0& 0& 0\\
			\hline
			\hline 
			Total &  & 436 &156 & 122 \\
			\hline
		\end{tabular}
	\end{table}

	\begin{table}[hbt!]
		\caption{\label{tab:fine_biv_consider_specificity} Bivariate LGCP estimates with specificity = 1 vs specificity $>$ 1.} 
		\scriptsize
		\begin{center}
			\begin{threeparttable}
				\vspace{-5mm}
				\begin{tabular}{llcccccc}
					\toprule
					\multicolumn{2}{c}{} &   \multicolumn{3}{c}{specificity=1}
					& \multicolumn{3}{c}{specificity$>$1}\\
					\cmidrule(lr){3-5}
					\cmidrule(lr){6-8}
					Parameter & &BH &FE &CP  
					&BH &FE &CP\\
					\cmidrule(r){1-1}\cmidrule(r){2-8}
					$\sigma$    
					&Lower  
					&1.31    &1.00    &2.34
					&1.48    &2.66    &1.02 \\
					&Median 
					&1.68    &1.02    &2.36
					&2.18    &3.16    &1.36 \\
					&Upper 
					&2.05    &1.07    &2.38
					&3.04    &3.99    &1.68 \\
					\\
					$\phi$   
					&Lower  
					&3.43    &12.59    &85.65
					&33.52    &68.83    &4.65 \\
					&Median 
					&6.74    &13.97    &135.80
					&64.04    &99.09    &8.93 \\
					&Upper 
					&13.06    &15.81    &212.33
					&114.98    &142.89    &16.50 \\
					\cmidrule{1-8}
					$\text{Intercept}$    
					&Lower  
					&-13.31    &-17.12  & 
					&-13.71    &-14.03  & \\
					&Median 
					&-11.45    &-15.87   &
					&-12.08    &-11.60   &\\
					&Upper 
					&-9.59    &-14.22   &
					&-10.23    &-9.29    &\\
					\\
					$ \log(\text{pop})$
					&Lower  
					&4.71    &0.13   &
					&2.23    &1.20   & \\
					&Median 
					&\textbf{5.44}    &\textbf{1.06}  &
					&\textbf{3.38}    &\textbf{2.36}  &\\
					&Upper 
					&6.29    &2.10    &
					&4.47    &3.67    &\\
					\\
					$ \text{elevation}$ 	
					&Lower  
					&-0.19    &-0.57    &
					&-1.70    &-1.37    &\\
					&Median 
					&0.92    &0.41    &
					&-0.13    &-0.20  &\\
					&Upper 
					&2.12    &1.30   & 
					&1.41    &0.85   & \\
					\\
					$\text{mdis}$    
					&Lower  
					&-5.39    &-3.94   &
					&-4.18    &-4.00   &    \\
					&Median 
					&\textbf{-3.45}    &-1.68  &  
					&\textbf{-2.02}    &-0.89  & \\
					&Upper 
					&-1.61    &0.57   & 
					&-0.12    &1.97   & \\
					\\
					$ \text{lon}$ 	
					&Lower  
					&1.23    &-2.01   & 
					&3.38    &0.05    & \\
					&Median 
					&\textbf{4.87}    &1.30    &
					&\textbf{6.47}    &\textbf{4.14}  &\\
					&Upper 
					&8.22    &4.76   &
					&10.75    &8.54  & \\
					\\
					$ \text{lat}$  
					&Lower  
					&-0.42    &-3.96  &
					&-0.18    &-3.58   &\\
					&Median 
					&2.03    &-1.54   &
					&2.26    &-0.22 &\\
					&Upper 
					&4.68    &0.99   &
					&4.87    &2.73   &\\
					\\
					$ \text{lon*lat}$ 
					&Lower  
					&-2.89    &-14.55  & 
					&-9.41    &-15.89  & \\
					&Median 
					&1.48    &\textbf{-7.15}  &
					&-2.71    &-6.82 & \\
					&Upper 
					&5.83    &-0.74  & 
					&3.28    &1.20  & \\
					\\
					\bottomrule
				\end{tabular}
				\begin{tablenotes}
					\scriptsize
					\item Note: Bold text indicates significant parameter estimates (at 95\% level)
				\end{tablenotes}
			\end{threeparttable}
		\end{center}
	\end{table}
	
	To see how this also affects model estimates, we fit LGCP models to two subsamples of the original data, one with only the specificity = 1 cases and another with specificity $>$ 1 observations. Table \ref{tab:fine_biv_consider_specificity} presents the parameter estimates from the two samples using a bivariate LGCP model. For the covariates, estimated coefficients are consistent overall between the two groups although there are some differences between coefficients regarding coordinates. However, the main difference between two subsamples is the covariance parameters of log intensity functions. Specifically, considering all $\hat{\sigma}$ and $\hat{\phi}$, we find that the case of specificity $>1$, have a smaller within-group clustering size and smaller cross-group correlation (in magnitude). The reason the specificity $>$ 1 sample indicates a smaller clustering size and smaller cross-correlation is because observations with little accuracy will be recorded as if occurred at identical locations, i.e., having smaller pairwise (with-group and cross-group) distances.\footnote{Note that in the FE sample the marginal correlation does not vary much (less than that of BH) between two subsamples considering specificity. This is consistent with the information in Table \ref{tab:spec}, i.e., for FE, two cases have similar number of points and similar mean of pairwise distances.} 
	
	Aside from the estimated spatial dependence, the two samples (specificity $= 1$ vs. specificity $>$ 1) have very different estimated variances of the log intensity process. For example, in the sample with specificity $>$ 1, $\hat{\sigma}^2_{FE}+\hat{\sigma}^2_{CP}$ is larger than those of the specificity $=$ 1 sample. This is because in the specificity $>$ 1 sample, events are inaccurately recorded and covariates may not provide enough information of intensity. As such, for FE, the realised $\Lambda(\mathbf{s})$ has more variation across its expectation explained by covariates. However, $\hat{\sigma}^2_{BH}+\hat{\sigma}^2_{CP}$ is smaller for the sample of specificity $>$ 1, because in this sample, the number of events with specificity $>$ 1 by BH is far less than that for specificity $=$ 1 (Table \ref{tab:spec}). Taken together, these results demonstrate the importance of accounting for the level of precision in supplied coordinate information, as the different samples analyzed here have both distinct cluster sizes and potential for meaningful clustering. 
	
	\begin{figure}[hbt!]
		\centering
		\begin{subfigure}{0.49\textwidth}
			\includegraphics[width=0.9\textwidth]{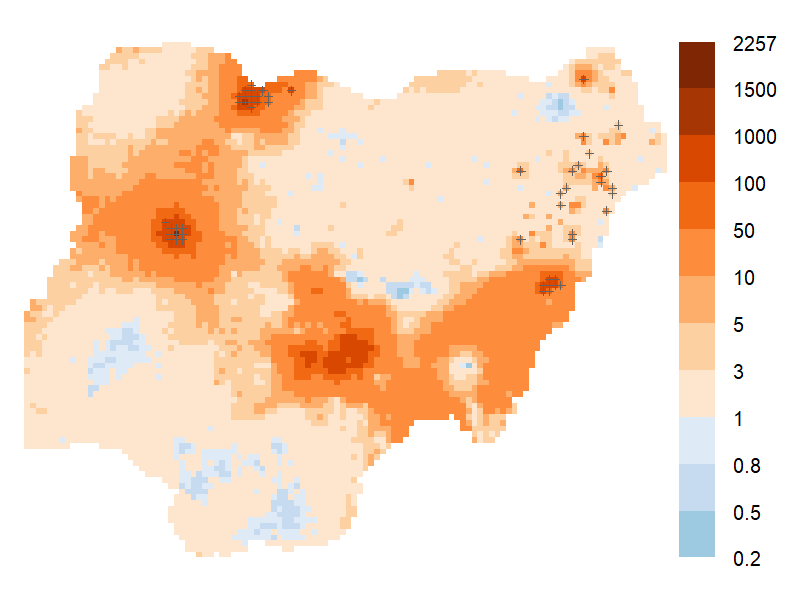}
			\caption{BH}
			\label{fig:biv_bb_median_ratio_contour_specif}
		\end{subfigure}
		\begin{subfigure}{0.49\textwidth}
			\includegraphics[width=0.9\textwidth]{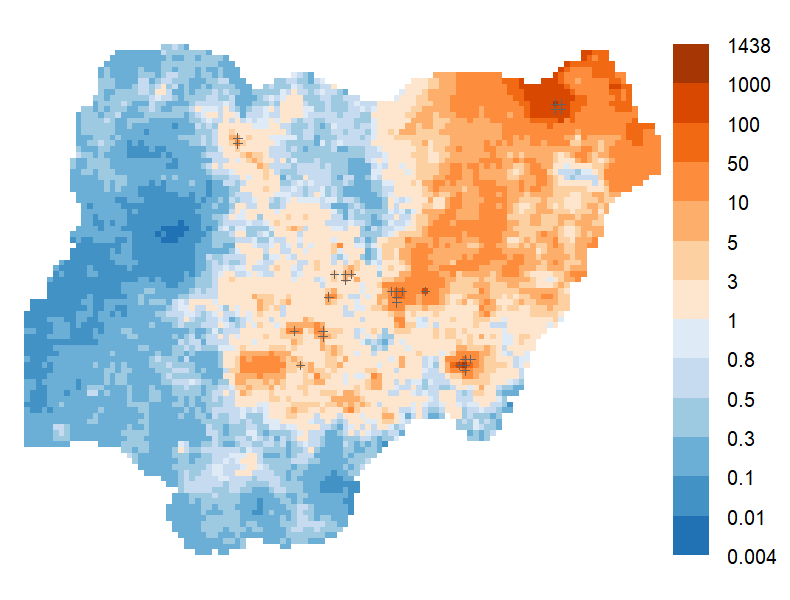}
			\caption{FE}
			\label{fig:biv_ff_median_ratio_contour_specif}
		\end{subfigure}
		\caption{Median ratio of intensity from bivariate LGCP analysis with all specificity vs specificity = 1. Locations where the ratio is significantly larger than 1 (at 1\% level) is marked with $+$ and those with ratio significantly smaller than 1 are marked with $\times$. }
		\label{fig:ratio-specif-biv}
	\end{figure}

	Figure \ref{fig:ratio-specif-biv} shows the ratio of estimated intensity for the two subsamples. Looking first to the BH results in panel (a), the median intensity ratio across space has a range $[0.2, 2257]$. The median and mean of these values are 2.62 and 10.06 respectively. We see two characteristics: 1) in almost all pixels (red regions), the estimated intensity is larger for all specificity, extremely red pixels are because of covariates effects (ex. coordinates and their interaction). 2) In few pixels, the estimated intensity for all specificity is smaller (but near to) that of specificity = 1. This is consistent with our intuition, as the subsample with all levels of specificity included should have larger intensity. Turning to the FE results, the median intensity ratio across space has a range $[0.004,1438]$. The median and mean of these values are 0.75 and 6.47 respectively. The estimated intensities for the all specificity sample in the center of Nigeria (light red regions where attacks happen) are about 1 to 3 times as large as that of specificity = 1.\footnote{Note that this is because the number attacks is twice as large in the ``all specificity'' sample.} In the remaining regions, the intensity is determined by the covariate longitude, i.e. western part is smaller for all specificity and eastern part is larger. For both groups, the ratio of intensities are not significantly different from 1 (at the significance level of 1\%), except very few pixels (extremely red regions where attacks occur and they are also influenced by coordinates covariates).

	\clearpage
	\newpage
	\section{\label{sec:uni_diff_case}Univariate LGCP analysis in different cases}
	
	\begin{table}[H]
		\caption{ Summary of univariate LGCP covariance function parameter estimates in different cases (fine cov vs coarse cov; fine cov vs Non-jittered; specificity=1 vs specificity$>$1). } 
		\scriptsize
		\begin{center}
			\begin{tabular}{llrrrrrrrrrr}
				\toprule
				\multicolumn{2}{c}{} &   
				\multicolumn{2}{c}{fine cov}
				& \multicolumn{2}{c}{coarse cov}
				& \multicolumn{2}{c}{non-jittered}
				& \multicolumn{2}{c}{specificity=1}
				& \multicolumn{2}{c}{specificity$>$1}
				\\
				\cmidrule(lr){3-4} 
				\cmidrule(lr){5-6}
				\cmidrule(lr){7-8}
				\cmidrule(lr){9-10}
				\cmidrule(lr){11-12}
				
				Parameter & &BH &FE   &BH &FE  &BH &FE &BH &FE &BH &FE\\
				\cmidrule(r){1-1}\cmidrule(r){2-12}
				$\sigma$    
				&Lower  
				&2.14    &2.45
				&2.41    &2.55
				&1.55    &2.28   
				&2.12    &2.14
				&1.73    &2.48
				\\
				&Median 
				&2.59    &3.10   
				&2.81    &3.22
				&1.91    &2.91    
				&2.58    &3.01
				&2.14    &3.23\\
				&Upper 
				&3.21    &4.11    
				&3.31    &4.18
				&2.44    &3.81     
				&3.23    &4.12
				&2.84    &4.48\\
				\\
				$\phi$   
				&Lower  
				&25.72    &50.72    
				&12.25    &48.03
				&32.33    &55.85     
				&20.88    &51.86
				&10.68    &45.79\\
				&Median 
				&38.07    &77.79   
				&17.61    &75.00
				&50.41    &87.05    
				&33.70    &93.96
				&18.20    &72.96\\
				&Upper 
				&59.34    &132.15
				&25.69    &123.66
				&81.95    &148.18    
				&54.30    &186.83
				&34.11    &131.45\\
				\\
				\bottomrule
			\end{tabular}
		\end{center}
		\label{tab:uni_different_cases_covariance_par}
	\end{table}

	\begin{table}[H]
		\caption{ Summary of univariate LGCP coefficient estimates in different cases (fine cov vs coarse cov; fine cov vs Non-jittered; specificity=1 vs specificity$>$1). Bold faced values denote significant estimates at 5\% level.} 
		\scriptsize
		\begin{center}
			\begin{tabular}{llrrrrrrrrrr}
				\toprule
				\multicolumn{2}{c}{} &   
				\multicolumn{2}{c}{fine cov}
				& \multicolumn{2}{c}{coarse cov}
				& \multicolumn{2}{c}{Non-jittered}
				& \multicolumn{2}{c}{specificity=1}
				& \multicolumn{2}{c}{specificity$>$1}
				\\
				\cmidrule(lr){3-4} 
				\cmidrule(lr){5-6}
				\cmidrule(lr){7-8}
				\cmidrule(lr){9-10}
				\cmidrule(lr){11-12}
				
				Parameter & &BH &FE   &BH &FE  &BH &FE &BH &FE &BH &FE\\
				\cmidrule(r){1-1}\cmidrule(r){2-12}
				$\text{Intercept}$    
				&Lower  
				&-12.19    &-10.54    
				&-10.44    &-10.40  
				&-11.96    &-10.89    
				&-13.26    &-11.59 
				&-14.07    &-11.46\\
				&Median 
				&-11.22    &-9.07 
				&-9.78     &-8.83
				&-11.10    &-9.31     
				&-12.22    &-9.82  
				&-12.78    &-9.75\\
				&Upper 
				&-9.94    &-6.39 
				&-8.95    &-6.06
				&-10.08    &-6.71     
				&-10.96    &-7.21   
				&-11.64    &-6.83\\
				\\
				$ \log(\text{pop})$
				&Lower  
				&4.50    &1.11  
				&0.37    &-0.42
				&3.40    &1.01      
				&4.72    &0.59     
				&2.06    &1.27\\
				&Median 
				&\textbf{5.26}    &\textbf{2.03}     
				&\textbf{1.33}           &1.55
				&\textbf{4.01}    &\textbf{1.86}     
				&\textbf{5.60}    &\textbf{1.56}    &\textbf{2.98}   &\textbf{2.35}\\
				&Upper 
				&6.14    &2.93      
				&2.29    &3.32
				&4.65    &2.73     
				&6.65    &2.69    
				&4.12    &3.58\\
				\\
				$ \text{elevation}$ 	
				&Lower  
				&-1.37    &-1.33     
				&-0.81    &-0.76
				&-0.96    &-1.22     
				&-1.32    &-1.27     
				&-1.27    &-1.28\\
				&Median 
				&-0.07    &-0.22     
				&0.37     &0.27
				&0.02    &-0.17  
				&0.04    &-0.04      
				&-0.11   &-0.02\\
				&Upper 
				&1.02    &0.78   
				&1.34    &1.24
				&0.92    &0.76     
				&1.24    &1.15   
				&1.04    &1.23\\
				\\
				$\text{mdis}$    
				&Lower  
				&-3.42    &-3.76    
				&-2.82    &-2.69
				&-3.16    &-3.54     
				&-4.33    &-4.34     
				&-3.83    &-4.05\\
				&Median 
				&\textbf{-1.70}    &-1.14 
				&\textbf{-1.69}             &-0.22
				&\textbf{-1.75}    &-1.03
				&\textbf{-2.36}    &-1.50     
				&\textbf{-2.43}   &-1.01
				\\
				&Upper 
				&-0.17    &1.52     
				&-0.57    &2.38
				&-0.19    &1.39      
				&-0.70    &1.14     
				&-0.94    &1.97\\
				\\
				$ \text{lon}$ 	
				&Lower  
				&3.17    &-0.05     
				&2.75    &-1.02
				&2.25    &-0.22    
				&3.97    &-2.26   
				&5.48   &0.41\\
				&Median 
				&\textbf{5.42}    &3.24    
				&\textbf{4.32}        &2.26
				&\textbf{4.24}    &3.09     
				&\textbf{6.25}    &2.05      
				&\textbf{8.20}   &\textbf{3.98}\\
				&Upper 
				&7.47    &7.60   
				&5.79    &6.38
				&6.00    &7.51    
				&8.79    &6.88   
				&10.61   &9.74\\
				\\
				$ \text{lat}$  
				&Lower  
				&0.67    &-3.07    
				&-0.26   &-2.83
				&0.26    &-2.88     
				&0.06    &-4.62    
				&-0.11   &-3.55\\
				&Median 
				&\textbf{2.20}    &-0.54     
				&1.02           &-0.36
				&\textbf{1.68}    &-0.45      
				&\textbf{1.73}    &-1.22   
				&1.64   &-0.63\\
				&Upper 
				&4.12    &2.03     
				&2.41    &2.29
				&3.41    &2.22     
				&3.73    &1.50      
				&4.04    &2.49\\
				\\
				$ \text{lon*lat}$ 
				&Lower  
				&-2.45    &-13.27   
				&-0.88    &-12.02
				&-1.95    &-13.00    
				&-1.89    &-18.87      
				&-9.15    &-16.65 \\
				&Median 
				&1.09    &-6.33    
				&1.97    &-4.89
				&1.06    &-5.84     
				&1.58    &-8.30    
				&\textbf{-4.90}   &-6.87\\
				&Upper 
				&4.73    &0.45    
				&5.05    &1.73
				&4.31    &1.02    
				&5.86    &0.09   
				&-0.43   &0.71\\
				\\
				\bottomrule
			\end{tabular}
		\end{center}
		\label{tab:uni_different_cases_coefficient}
	\end{table}

\end{appendices}

\end{document}